\documentclass[multi]{cambridge6A}

\usepackage{amsfonts,amsmath,amssymb}

\usepackage[numbers,merge]{natbib}
\usepackage{rotating}
\usepackage{floatpag}
\rotfloatpagestyle{empty}
\usepackage{amsthm}
\usepackage{graphicx}

\begin{document}

\alphafootnotes
\author{Lev Pitaevskii\footnotemark and Sandro Stringari\footnotemark}
\chapter{Second Sound in Ultracold Atomic Gases}

\footnotetext[1]{INO-CNR BEC Center and Dipartimento di Fisica, 
Universit\`a di Trento, Via Sommarive 14, I-38123 Povo, Italy; 
Kapitza Institute for Physical Problems, RAS, 
ul Kosygina 2, 119334 Moscow, Russia} 
\footnotetext[2]{INO-CNR BEC Center and Dipartimento di Fisica, 
Universit\`a di Trento, Via Sommarive 14, I-38123 Povo, Italy}

\begin{abstract}
We  provide an  overview of  the recent  theoretical and  experimental
advances  in the  study of  second  sound in  ultracold atomic  gases.
Starting from the Landau's two fluid hydrodynamic equations we develop
the  theory  of  first  and second  sound  in  various  configurations
characterized by  different geometries and quantum  statistics.  These
include the weakly  interacting 3D Bose gas,  the strongly interacting
Fermi gas at  unitarity in the presence of highly  elongated traps and
the     dilute    2D     Bose     gas,     characterized    by     the
Berezinsky-Kosterlitz-Thouless transition. An explicit comparison with
the propagation  of second sound  in liquid  Helium is carried  out to
elucidate  the  main  analogies  and differences.   We  also  make  an
explicit comparison with the available experimental data and point out
the crucial  role played by  the superfluid density in  determining the
temperature dependence of the second sound speed.
\end{abstract}

\section{Introduction}

Superfluidity is  one of the most  challenging features characterizing
the  behavior of  ultra cold  atomic gases  and provides  an important
interdisciplinary  connection  with  the physics  of  other  many-body
systems    both    in    condensed   matter    (superfluid    Helium,
superconductivity) and  high energy physics  (nuclear superfluidity, 
neutron stars).  It  is a
phenomenon deeply related to Bose-Einstein condensation, which, in the
case of Fermi superfluids, corresponds to the condensation of pairs.
Several  manifestations  of  superfluidity  have  already  emerged  in
experiments on  ultracold atomic gases,  either in the case  of bosons
and fermions  \cite{PitaevskiiStringari}. 

Important phenomena  are the
consequence  of   the  irrotational  nature  of   the  velocity  field
associated  with the  superfluid flow. They include 
 the occurrence  of
quantized vortices, which is actually a direct manifestation of the
absence of shear viscosity,
 the quenching  of the moment  of inertia,  and
the
behavior of the collective oscillations at zero temperature. 
Vortices
in  a  superfluid  are  characterized   by  the  quantization  of  the
circulation  $\oint {\bf  v}_{s}d{\bf l}$  of the  superfluid velocity
around the vortex line.  They have become experimentally
available using  different approaches.  A first  procedure consists of
creating   the  vortical   configuration  with   optical  methods   in
two-component condensates \cite{Matthews1999}.  A second method, which
shares a  closer resemblance  with the  rotating bucket  experiment of
superfluid helium, makes use of  a suitable rotating modulation of the
trap  to stir  the  condensate \cite{Madison2000}.   Above a  critical
angular  velocity one  observes the  formation of  vortices which  are
imaged after  expansion. Configurations  containing a large  number of
vortices  can  be  created  by  stirring  the  condensate  at  angular
velocities close  to the centrifugal  limit and have been  realized in
both     Bose    \cite{AboShaeer2001,Coddington2003},     and    Fermi
\cite{Zwierlein2005} superfluid gases.  
 The quenching of the moment of
inertia due to superfluidity shows up  in the peculiar behavior of the
scissors  mode \cite{GueryOdelin1999,Marago2000}  and its  temperature
dependence  has  been  detected  directly  by  measuring  the  angular
momentum of a trapped superfluid  rotating at a given angular velocity
\cite{Riedl2011}.   The  effects of  irrotationality on  the collective
oscillations of  both Bose  and Fermi superfluid  gases have  been the
object  of  extensive  theoretical  and  experimental  work,  and  are
accurately described by superfluid hydrodynamics in the
presence of  harmonic traps \cite{Stringari1996}. 
  
An important consequence of  superfluidity  
is the occurrence  of a  critical velocity  below which  the
superfluid can  move without dissipation (Landau's  criterion).  These
experiments  were   carried  out  with  moving   focused  laser  beams
(\cite{Onofrio2000}) as  well as  with moving  one-dimensional optical
lattices \cite{Miller2007}. Other  manifestations of superfluidity are
provided by the coherent  macroscopic behavior responsible for quantum
tunneling in  the double well  geometry \cite{Albiez2005} and,  in the
case  of  Fermi  superfluids,  by  the appearance  of  a  gap  in  the
single-particle  excitation  spectrum,  revealed  by  radio  frequency
transitions  \cite{Chin2004}.    Let  us   also  mention   the  recent
experimental  investigation of  the equation  of state  of a  strongly
interacting Fermi gas \cite{Ku2012}. This measurement has revealed the
typical  $\lambda$-behavior of  the  specific heat  at the  superfluid
transition and an accurate determination of the critical temperature.

A key property of a superfluid at finite temperature is its
two--fluid nature, i.e.~ the simultaneous presence of two flows:
a dissipationless superfluid flow and a normal flow.
The propagation of second sound \cite{Donnelly2009},
which represents the topic of discussion
of the present work, is a direct manifestation of this two--fluid nature.
An  important feature  of second  sound is
that it can  provide unique information on  the temperature dependence
of the superfluid density, as proven in the case of superfluid helium.
For  this  reason its  investigation  permits  to better  clarify  the
conceptual   difference   between  superfluidity   and   Bose-Einstein
condensation. This  distinction is particularly important  in strongly
interacting gases  (like the Fermi  gas at  unitarity) and it  is even
more  dramatic in  two  dimensions where  long  range order,  yielding
Bose-Einstein condensation, is ruled out  at finite temperature by the
Hohenberg-Mermin-Wagner theorem.

In  this paper  we summarize  some  advances in  the understanding  of
second sound  in quantum gases  that have been  achieved theoretically
and experimentally in the last few  years. In Section 2 we discuss the
general behavior of the Landau's  equations of two fluid hydrodynamics.
In Section 3  we discuss the solutions of these  equations in the case
of a 3D  dilute Bose gas. In  Section 4 we apply the  equations to the
case  of a  Fermi gas  at unitarity  in the  presence of  tight radial
trapping  conditions giving  rise to  highly elongated  configurations
which  are  suitable  for  the  experimental  investigation  of  sound
waves. Finally in Section 5 we discuss the behavior of second sound in
2D  Bose  gases where its measurement could  provide
unique information on the behavior of the superfluid density.

\section{Two-fluid hydrodynamics: first and second sound}

In the present work  we will describe the macroscopic dynamic  behavior of a
superfluid at  finite temperature.  We will  consider situations where
the free  path of the elementary  excitation is small compared  to the
wavelength of  the sound wave  and local thermodynamic  equilibrium is
ensured  by  collisions.   This  condition is  rather  severe  at  low
temperature where collisions  are rare.  It permits  to define locally
(in space  and time) the  thermodynamic properties of the  fluid, like
the temperature,  the pressure etc..  The general system  of equations
describing the macroscopic dynamic behavior of the superfluid at finite
temperature  was  obtained  by   Landau  \cite{Landau1941}  (see  also
Ref.~\cite{LandauLifshitz1987}).

In the  following we will  limit ourselves  to the description  of 
small velocities and small amplitude oscillations, corresponding to the 
linearized solutions of the equations of motion.  The propagation of
such oscillations at finite temperature  provides an important tool to
investigate the  consequences of superfluidity.  The  equation for the
superfluid velocity  ${\bf v}_{s}$ obeys the  fundamental irrotational
form
\begin{equation}
m\frac{\partial {\bf v}_{s}}{\partial t}+\nabla (\mu+V_{ext}) =0  
\label{stringari_dvsdt}
\end{equation}
fixed  by  the chemical  potential  $\mu(\rho,T)$  of uniform  matter
which, in  general, depends on the  local value of the  density and of
the temperature.

The equation  for the density $n$ has  the usual  form of  the continuity
equation:
\begin{equation}
\frac{\partial n }{\partial t}+{\rm div}{\bf j}=0.  
\label{stringari_con2}
\end{equation}
where the current density  ${\bf j}$ can be separated into  the normal and
superfluid components ${\bf j}=n_{s}{\bf v}_{s}+n_{n}{\bf v}_{n}$  
with $n_{s}+n_{n}=n$. The  time derivative of  the current
density is  equal to the  force per unit  volume which, in  the linear
approximation, is  fixed by the  gradient of  the pressure and  by the
external potential (Euler equation):
\begin{equation}
\frac{\partial {\bf j}}{\partial t}
+\frac{\nabla P}{m} + \frac{n}{m} \nabla V_{ext}=0.  
\label{stringari_djdt}
\end{equation}
The last  equation is the equation  for the entropy density  $s$. This
can  be  derived  from  general  arguments.  In  fact  if  dissipative
processes  such as  viscosity and  thermoconductivity are  absent, the
entropy is  conserved and  the equation  for $s$ takes  the form  of a
continuity   equation.   Furthermore   only   elementary   excitations
contribute to the entropy whose transport is hence fixed by the normal
velocity ${\bf v}_{n}$ of the fluid. The equation for the entropy then
takes the form
\begin{equation}
\frac{\partial s}{\partial t}+{\rm div}(s {\bf v}_{n})=0  
\label{stringari_ds}
\end{equation}
where,  in the  linear regime,  the entropy  entering the  second term
should be calculated  at equilibrium.

The  thermodynamic quantities  entering  the above  equations are  not
independent and obey the Gibbs-Duhem thermodynamic equation 
$n d\mu =-sdT +dP$.
It is worth mentioning that  also this thermodynamic identity is valid
only  in the  linear  regime.  In fact  in  general the  thermodynamic
functions can exhibit a dependence also on the relative velocity ${\bf
v}_{n}-{\bf v}_{s}$ between the normal and the superfluid velocities.

At $T=0$  where $n_n=0$, the  entropy identically vanishes  and the
Euler  equation  (\ref{stringari_djdt}),  thanks  to  the  Gibbs-Duhem
relation,     coincides    with     equation
(\ref{stringari_dvsdt}). The two-fluid   hydrodynamic   equations
(\ref{stringari_dvsdt}-\ref{stringari_ds}) then reduce to the equation
of  continuity and  to the  equation for  the irrotational  superfluid
velocity. Viceversa, for temperatures  larger than the critical value,
where  the  superfluid  density  vanishes and  the  equation  for  the
superfluid velocity can  be ignored, the other  three equations reduce
to the classical equations of dissipationless hydrodynamics.

Let  us for  the moment  ignore the  external potential  $V_{ext}$. By
looking  for plane  wave  solutions  varying in  space  and time  like
$e^{-i\omega(t-x/c)}$,  after a lengthy but straightforward calculation 
one obtains (see, for
example, \cite{PitaevskiiStringari}), the Landau's equation  for the sound
velocity \cite{Landau1941}.
\begin{equation}
c^{4}
-\left[ \frac{1}{mn\kappa_s}
+\frac{n_{s}T \bar{s}^{2}}{m n_{n}\bar{c}_{v}}\right] c^{2}
+\frac{n_{s}T \bar{s}^{2}}{m n_{n}\bar{c}_{v}}\frac{1}{mn\kappa_T} =0\ ,  
\label{stringari_c4}
\end{equation}
where $\bar{c}_{v}  = T(\partial \bar{s}/\partial T)_n$  is the
specific heat  at constant  volume per particle, 
$\kappa_s=(\partial n/\partial  P)_{\bar{s}}/n$  
and $\kappa_T=(\partial n/\partial  P)_T/n$
are, respectively, the adiabatic  and isothermal compressibilities while
$\bar{s}=s/n$ is the entropy per particle. 

\begin{figure}
  \begin{minipage}{.47\textwidth}
    \centering
    \includegraphics[width=\textwidth]
    {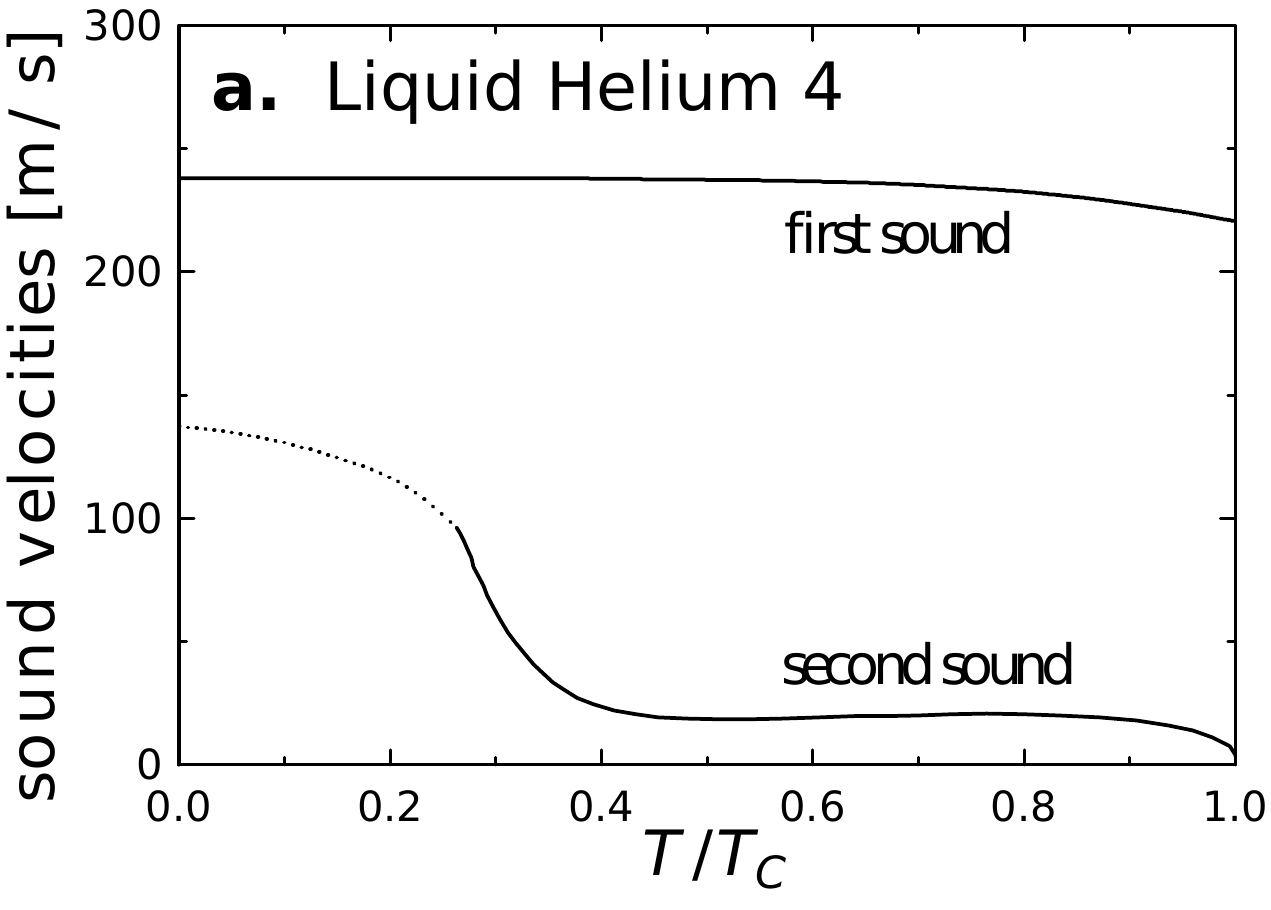}
  \end{minipage}
  \hspace*{.04\textwidth}
  \begin{minipage}{.47\textwidth}
    \centering
    \includegraphics[width=\textwidth]
    {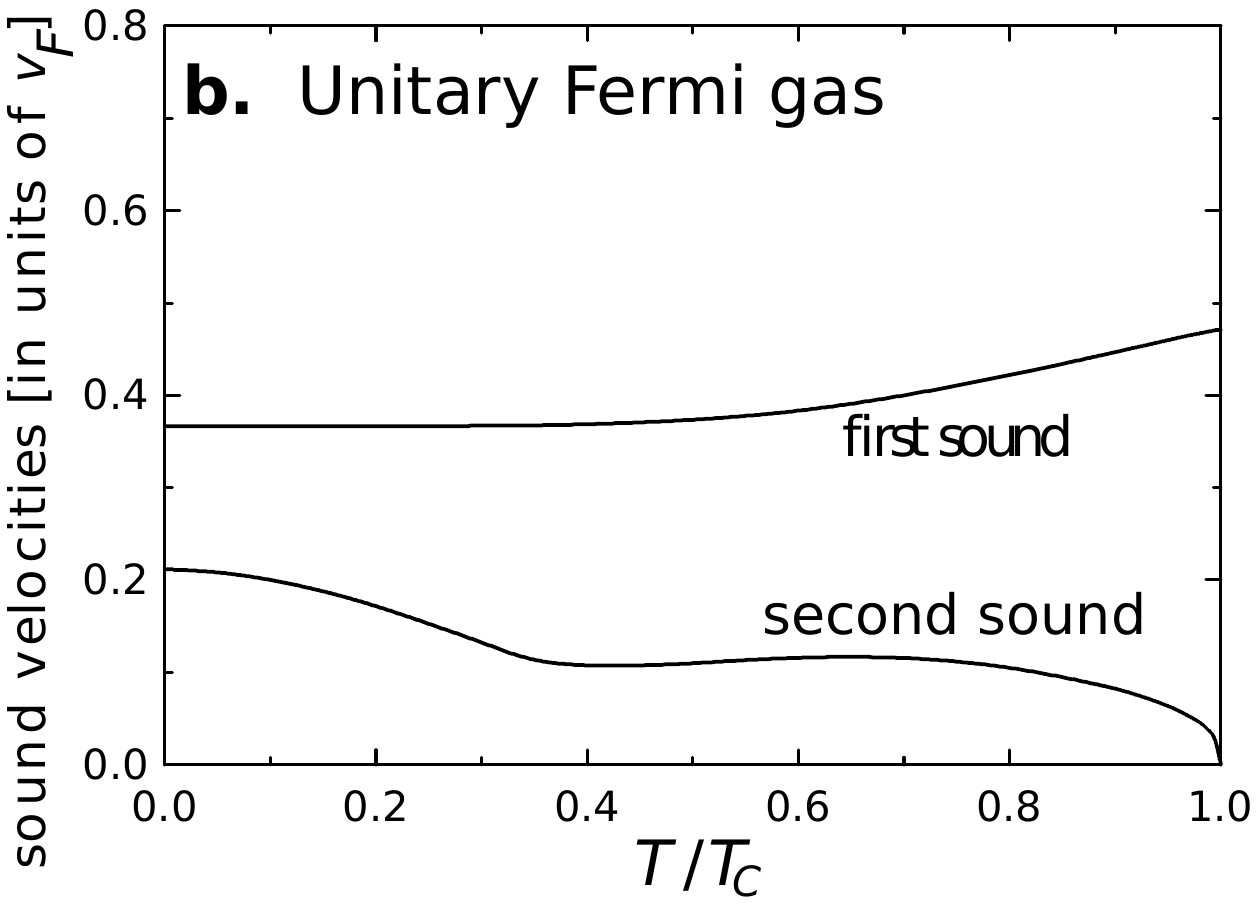}
  \end{minipage}
  \caption{ \label{stringari_He4_TaylorFig} 
\textbf{(a)} 
Experimental values of first and second sound speeds in superfluid
${}^4\mathrm{He}$.   Dots correspond  to  the theoretical  calculation
accounting  for the contribution  of the  phonon-roton excitations  to the
thermodynamic functions. 
\textbf{(b)}
Two-fluid sound speeds in a uniform Fermi gas at unitarity calculated 
using the NSR thermodynamic functions. From \cite{Taylor2009}
}
\end{figure}

If  $n_s \ne  0$ equation  (\ref{stringari_c4}) gives  rise to  two
distinct sound velocities,  known as first and second  sound.  This is
the consequence of the fact that in a superfluid there are two degrees
of freedom associated with the  normal and superfluid components.  The
existence of  two types  of sound waves  in a  Bose-Einstein condensed
system was  first noted by  Tisza \cite{Tisza1940}.  The  second sound
velocity   was    measured   in    superfluid   $^4$He    by   Peshkov
\cite{Peshkov1946}.  The values of  the two sound velocities, measured
in  superfluid $^4$He,  are reported  in 
Fig.~\ref{stringari_He4_TaylorFig}(a)  
as a
function of $T$.

The knowledge of the two sound  velocities can be used to evaluate the
density  strengths associated  with  the corresponding  modes and  the
consequent possibility  of exciting  them with  a density  probe.  For
example, using a sudden laser perturbation, one excites both first and
second sound with a relative weight given by the relative contribution
of the two modes to the inverse energy weighted moment
\begin{equation} 
\int_{-\infty}^\infty    d\omega \frac{1}{\omega}   S(\mathbf{q},\omega)
= \frac{1}{2} n\kappa_T
\label{stringari_kappaT1}
\end{equation}
also      know      as      the     compressibility      sum      rule
(\cite{PitaevskiiStringari}~\S   7),    fixed   by    the   isothermal
compressibility  and  here derived  in  the  limit of  small  momentum
transfer      $\mathbf{q}$.       In Eq.~(\ref{stringari_kappaT1}),
$S(\mathbf{q},\omega)$ is the dynamical structure factor with momentum
$\mathbf{q}$ and frequency $\omega$. Taking into account the fact that
at small wave  vectors not only the inverse energy  weighted sum rule,
but  also the  energy  weighted  moment $\int_{-\infty}^\infty  \omega
S(\mathbf{q},\omega)  =   q^2/2m$, known as $f$-sum rule, is exhausted 
 by   the  two  sound
modes~\cite{NozieresPines},  one  straightforwardly   finds  that  the
relative contributions to the compressibility sum rule 
(\ref{stringari_kappaT1})
from each sound
are given by~\cite{Hu2010}
\begin{align}
	W_1 &\equiv \frac{1 - mn\kappa_T c_2^2}{2m(c_1^2-c_2^2)},
	&
	W_2 &\equiv \frac{mn\kappa_T c_1^2 - 1}{2m(c_1^2-c_2^2)}.
	\label{stringari_Hu}
\end{align}

The Landau's equation (\ref{stringari_c4}) is  easily solved as $T \to
T_c$, i.e.   close to the  transition temperature where  $\rho _{s}\to
0$. In  this case  the upper   (first sound) and lower  (second sound) solutions are given  by 
\begin{align}
c^{2}_1 &=\frac{1}{mn\kappa_s}
=\frac{1}{m}
\left(\frac{\partial P}{\partial n}\right)_{\bar{s}} \; ,
&
c_{2}^{2} &= \frac{n_{s}T\bar{s}^{2}}{mn_{n}\bar{c}_p} \; .
\label{stringari_firstsoundvelocity}
\end{align}
The first sound velocity is given by the isoentropic 
value and exhibits a continuous transition to the usual 
sound velocity above
$T_c$. The second sound velocity is instead 
fixed by the superfluid density  
and  vanishes  at the  transition where  $n_s=0$ in  3D
systems.  In the  above equation we have introduced  the specific heat
at constant  pressure $\bar{c}_p$, 
related  to $\bar{c}_v$ by  the thermodynamic
relation $\bar{c}_p/\bar{c}_v=  \kappa_T/\kappa_{{s}}$. 

Results  (\ref{stringari_firstsoundvelocity})   hold
also   in  the   $T\to  0$   limit  where   one  can   set  
$\kappa_T=\kappa_{\bar{s}}$ 
 $\kappa_T=\kappa_{{s}}$
and $\bar{c}_p=\bar{c}_v$.   In this  limit all
the thermodynamic functions as well as the normal density $\rho_n$ are
fixed by  the thermal excitation of  phonons and one can  easily prove
that Eq.  (\ref{stringari_firstsoundvelocity}) yields result 
$c_{2}^{2}=c^{2}_1/3$ for
the second sound velocity.

For  systems characterized  by a  small thermal  expansion coefficient
$\alpha=-(1/n)\partial n/\partial T|_P=(\kappa_T/\kappa_{{s}}-1)/T$, 
and  hence very  close values
of    the   isothermal    and    adiabatic   compressibilities),    Eq.
(\ref{stringari_firstsoundvelocity}) provides an excellent approximation to the second
sound  velocity  in  the  entire  superfluid  region  $0<T<T_c$.  More
precisely, the applicability            of             results
(\ref{stringari_firstsoundvelocity})  requires 
the condition
\begin{equation}
\frac{c^2_2}{c^2_1}\alpha T \ll 1 \; .
\label{stringari_condition}
\end{equation}
Systems satisfying the
condition  (\ref{stringari_condition})  include   not  only  the  well
celebrated superfluid  $^4$He but  also the  interacting Fermi  gas at
unitarity  (see  Section  4).    Furthermore,  since  in  these  cases
$c^2_1\sim (mn\kappa_s)^{-1} \sim (mn\kappa_T)^{-1}$, the second sound
contribution to the compressibility sum rule (\ref{stringari_kappaT1})
is negligible and consequently second  sound should be excited using a
thermal  rather  than   a  density  perturbation  \cite{Lifshitz1944}.
Equation~(\ref{stringari_firstsoundvelocity}) for $c_2$
points out explicitly the  crucial role
played  by  the  superfluid  density  and  was  actually  employed  to
determine  the temperature  dependence  of the  superfluid density  in
liquid $^4$He in a wide interval of temperatures  \cite{Dash1957}.

Equations  (\ref{stringari_firstsoundvelocity}) are
instead inadequate  to describe  the sound  velocities in  dilute Bose
gases  at intermediate  values  of  $T$, due  to  the high  isothermal
compressibility  exhibited by  these  systems as  we  will discuss  in
Sections 3 and 5.

\section{Second sound in  weakly interacting 3D Bose gases}

An important  consequence of  the high compressibilility  exhibited by
weakly interacting  Bose gases  is the  occurrence of  a hybridization
phenomenon  between first  and second  sound.  This  phenomenon, first
investigated in the seminal paper  by Lee and Yang \cite{LeeYang1959},
and  later discussed  by  Griffin and  co-workers  (see, for  example,
\cite{Zaremba1999,GriffinNikuniZaremba2009, Hu2010}) has been recently
investigated  in   detail  in   \cite{Verney2015}  where   a  suitable
perturbative approach  has been developed  to explore the  behavior of
second sound below and above the hybridization point. The mechanism of
hybridization is  caused by the  tendency of  the velocity of  the two
modes to  cross at  very low  temperatures, of the  order of  the zero
temperature  value of  the chemical  potential $\mu(T=0)  = gn$  where
$g=4\pi\hbar^2a/m$ is  the bosonic coupling constant.  This phenomenon
characterizes dilute Bose gases and  is absent in strongly interacting
superfluids. It  is natural to  call the  upper and lower  branches as
first  and second  sounds,  respectively. For  temperatures below  the
hybridization point  the velocity of  the upper branch  approaches the
Bogoliubov   value  $c_{B}=\sqrt{gn/m}$,   while   the  lower   branch
approaches  the Landau's  result  $c_{2}=c_B/\sqrt{3}$.   Above the hybridization point the role of first and second
sound is inverted, in the sense  that the lower (second sound) branch is
essentially an oscillation of the superfluid density which practically
coincides with the condensate density $n_0(T)$.

\begin{figure}
  \centering
  \includegraphics[width=.6\textwidth]
  {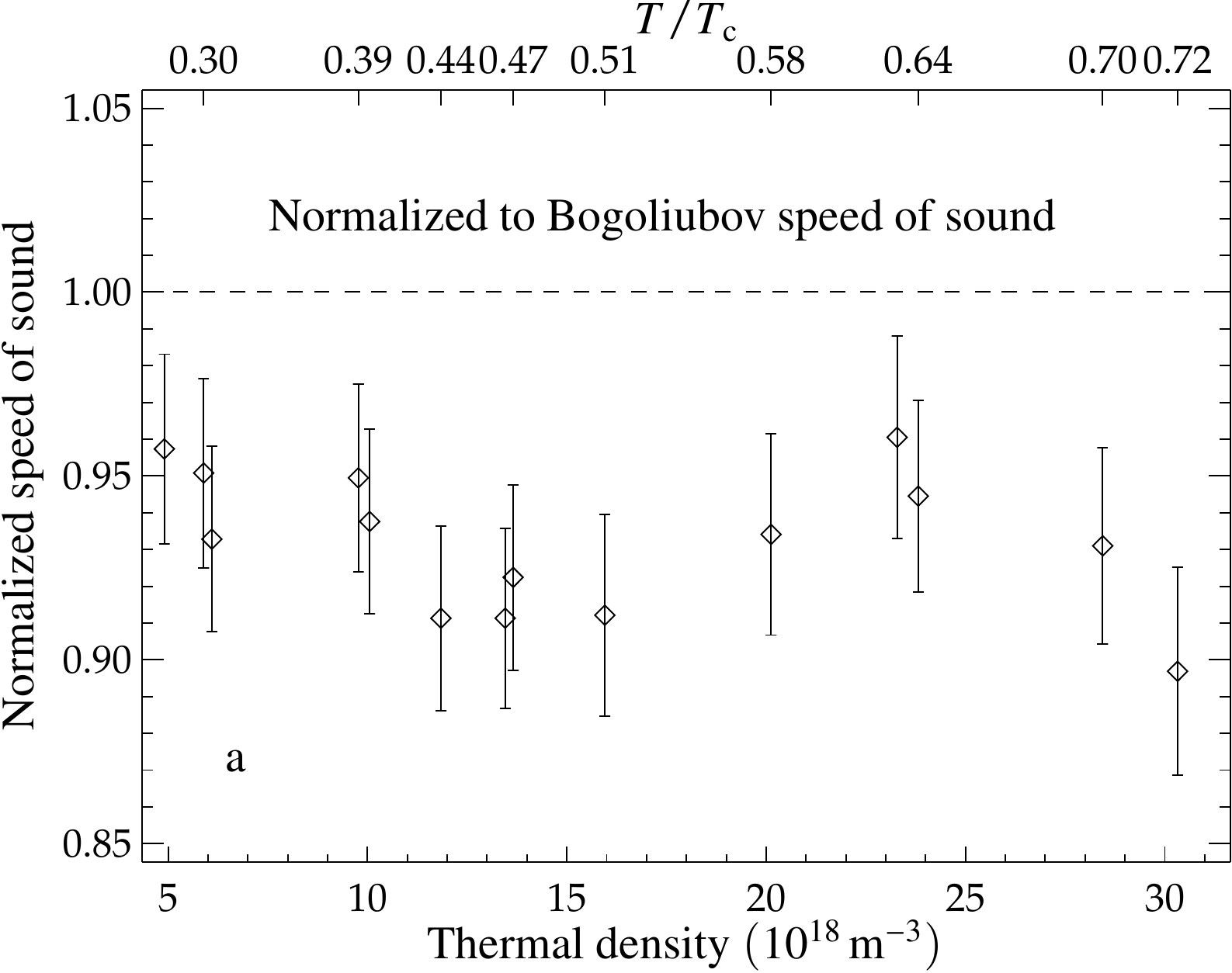}
  \caption{ \label{stringari_straten}
  Speed  of  sound  as  a  function  of  the  thermal
density. The  upper axis gives the  reduced temperature $T/T_\text{c}$
for the corresponding data point. The speed of sound is
normalized  to   the Bogoliubov sound velocity $\sqrt{gn_0(T)/2m}$
  based   on  the central
 BEC density. From \cite{Meppelink2009}.
  }
\end{figure}

For temperatures  higher than the hybridization  temperature ($k_BT\gg
gn$) a useful expression for the  second sound velocity is obtained by
evaluating   all  the   quantities  entering   the  quartic   equation
(\ref{stringari_c4}),  except the  isothermal compressibility  and the
superfluid density, using the ideal  Bose gas model.  One can actually
show that the adiabatic compressibility, the specific heat at constant
volume  and the  entropy  density  of a  weakly  interacting Bose  gas
deviate very  little from  the ideal  Bose gas  predictions in  a wide
interval  of   temperatures  above  the  hybridization   point.   This
simplifies     significantly  the solution     of     Equation
(\ref{stringari_c4}). In  fact in the ideal Bose gas model  one finds:
\begin{equation}
\frac{1}{mn\kappa_s}
+\frac{n_{s}T \bar{s}^{2}}{m n_{n}\bar{c}_{v}}
= \frac{n T \bar{s}^2}{m n_n\bar{c}_v}
\label{stringari_thermoIBG}
\end{equation}
for the coefficient of  the $c^2$ term.  It is now  easy to derive the
two solutions satisfying  the condition $c_{1}\gg c_{2}$  .  To obtain
the larger velocity  $c_{1}$ one can neglect the last  term in $\left(
\ref{stringari_c4} \right) $. Using the thermodynamic relations of the
ideal  Bose gas  model and  identifying  the normal  density with  the
thermal density ($n_T=n-n_0$) one obtains the prediction
(here and in the following we take the Boltzmann constant 
$k_{\mathrm{B}}=1$)
\begin{equation}
c_{1}^{2}=\frac{5}{3}\frac{g_{5/2}}{g_{3/2}} \frac{T}{m}  
\label{stringari_c1B}
\end{equation}
for the  first sound velocity \cite{LeeYang1959},  where $g_{5/2}$ and
$g_{3/2}$ are integrals  of the Bose distribution  function (see, for
example, \cite{PitaevskiiStringari}  \S 3.2).  To  calculate $c_{2}$
we   must  instead   neglect  the   $c^{4}$  term   in  $\left(   \ref
{stringari_c4}\right) $  and using  result (\ref{stringari_thermoIBG})
one finds the useful result
\begin{equation}
    c^2_2 = \frac{n_s}{n}\frac{1}{mn\kappa_T}
    \label{stringari_c-chiT}
\end{equation}
revealing   that   the   superfluid   density   and   the   isothermal
compressibility are  the crucial  parameters determining the  value of
the  second  sound  velocity  of  weakly-interacting  Bose  gases  for
$T\gg gn$. Furthermore in a weakly  interacting 3D Bose gas one can
safely identify (except close to $T_c$) the inverse  isothermal 
compressibility with its $T=0$
value $(n\kappa_T)^{-1}=gn$ yielding the simple expression
\begin{equation}
c^2_2= \frac{gn_s(T)}{m}
\label{stringari_simple}
\end{equation}
for  the second  sound velocity,  revealing that  second sound  can be
regarded,  in   this  temperature  range,  as   a  finite  temperature
generalization of  the Bogoliubov sound  mode propagating at  $T=0$ at
the velocity $\sqrt{gn/m}$ (at $T=0$ one has $n_s=n$).  Since
in a  weakly interacting  3D Bose  gas the  superfluid density  can be
safely approximated with  the condensate density 
($n_s(T)=n_0(T)$)
the above result also permits to understand that in the same regime of
temperatures  second  sound  corresponds  to  an  oscillation  of  the
condensate which can be easily  excited through a density perturbation
of the  gas and subsequently imaged.   In Fig.~\ref{stringari_straten}
we  report the  sound velocity  measured  in the  experiment of  \cite
{Meppelink2009}, where  a density perturbation  was applied to  a Bose
gas confined in  a highly elongated trap.  Due to  the harmonic radial
confinement the density $n$ entering  the Bogoliubov formula should be
actually  replaced by  $n/2$ where  $n$ is  the central  axial density
\cite{Zaremba1998} (a similar renormalization  is expected to occur
for the superfluid density and the condensate density at 
finite temperature).  
The good agreement
between  the   measured  sound  velocity  and   the  Bogoliubov  value
$\sqrt{gn_0(T)/2m}$, with the condensate fraction measured at the same
value  of  $T$,   permits  to  conclude  that   the  sound  excitation
investigated  in  \cite{Meppelink2009}  actually  corresponds  to  the
second sound mode described above.  The remaining small deviations are
likely   due   to   the   approximations   made   to   derive   result
(\ref{stringari_simple}) and to the identification $n_s=n_0$.

In order to explore the behavior  of the sound velocities in the whole
range of temperatures, including the  hybridization region at low $T$,
a more accurate knowledge of  the thermodynamic functions entering the
two-fluid hydrodynamic equations is needed. At low temperature ($T \ll
T_c$) one can  use Bogoliubov theory where  the elementary excitations
of the gas, whose dispersion is given by the  Bogoliubov law 
$\varepsilon(\mathbf{p})
=\sqrt{(p^2/2m) (p^2/2m+2gn)}$
are    thermally   excited    according    to    the   bosonic    rule
${N_\mathbf{p}(\varepsilon)=(e^{\varepsilon(\mathbf{p})/kT}-1)^{-1}}$.
Bogoliubov theory  fails at temperatures  of the order of  the critical
temperature  where  a  more  reliable  approach  is  provided  by  the
perturbation theory  developed in Ref.~\cite{Capogrosso2010}  based on
the   Beliaev    diagrammatic   technique   at    finite   temperature
\cite{Abrikosov1975}.  In this latter  approach thermal effects in the
Bogoliubov   excitation  spectrum   are   accounted   for  through   a
self-consistent procedure. At low temperatures this approach coincides
with  Bogoliubov   theory,  except   for  temperatures   smaller  than
$(na^3)^{1/4}gn$, \emph{i.e.}   at temperatures much smaller  than the
hybridization   point.   At   higher   temperatures   the  theory   of
Ref.~\cite{Capogrosso2010}  turns out  to be  very accurate  in dilute
gases when  compared with  exact Monte-Carlo simulations.   It follows
that,  at  least  for  small  values  of  the  gas  parameter  $na^3$,
Bogoliubov    theory     and    the    diagrammatic     approach    of
Ref.~\cite{Capogrosso2010} match  exactly in the  hybridization region
of temperatures  $T \sim gn$  and that the thermodynamic  behavior of
the gas is consequently under  control for all ranges of temperatures,
both below and above $gn$.

The coefficients  of the quartic equation  (\ref{stringari_c4}) depend
not only on  the equilibrium thermodynamic functions, but  also on the
normal and superfluid densities. The  normal density can be calculated
using the Landau's prescription
\begin{equation}
    m n_n = -
    \frac{1}{3} 
    \int \frac{d N_\mathbf{p}(\varepsilon)}{d \varepsilon} p^2 
    \frac{d \mathbf{p}}{(2\pi \hbar)^3}
  \label{stringari_Landau}
\end{equation}
in  terms of  the  elementary  excitations of  the  gas. The  Landau's
prescription  (\ref{stringari_Landau})   ignores  interaction  effects
among elementary excitations  and in a dilute Bose gas  is expected to
be  very accurate  in the  low temperature  regime characterizing  the
hybridization point  and to hold  also at higher  temperatures, except
close to the critical point.

A peculiar property of the weakly-interacting Bose gas is that all the
thermodynamic functions entering  Eq.~(\ref{stringari_c4}), as well as
the normal  density $\rho_n$, can be  written in a rescaled  form as a
function of the reduced temperature $\tilde{t}\equiv T/gn$ and of the
reduced chemical potential $\eta \equiv gn/T^0_C$ where
\begin{equation}
T^0_c=\frac{2\pi\hbar^2}{m}\left(\frac{n}{\zeta(3/2)}\right)^{2/3}
 \label{stringari_Tc0}
\end{equation}
is   the   critical   temperature   of  the   ideal   Bose   gas.   In
weakly-interacting  Bose  gases $T^0_c$  does  not  coincide with  the
actual critical temperature which contains a small correction fixed by
the value of the gas parameter $na^3$: $T_c=T_c^0 (1+ \gamma (na^3)^{1/3})$
with  $\gamma  \sim  1.3$  \cite{Baym1999,Arnold2001,Kashurnikov2001}.
The reduced chemical  potential can be also expressed in  terms of the
gas parameter as $\eta=2\zeta(3/2)^{2/3}(na^3)^{1/3}$.

Using Bogoliubov theory, the free energy $F=U-TS$ can be written as
\begin{equation}
    \begin{aligned}
        \frac{F}{gn N}=
        &\frac{1}{2}\left[1 + \frac{128}{15\sqrt{\pi}}(na^3)^{1/2}\right]\\
        &+ \frac{2\tilde{t}}{\zeta(3/2)\sqrt{2\pi}}\eta^{3/2} 
        \int_0^{\infty} \tilde{p}^2 
        \ln\left(1-e^{-\frac{\tilde{p}}{2\tilde{t}}\sqrt{\tilde{p}^2 + 4}}\right) 
        d \tilde{p}.\\
        \label{stringari_F}
    \end{aligned}
\end{equation}
where we have defined  $\tilde{p}=p/\sqrt{mgn}$ and, for completeness,
we have included  the Lee-Huang-Yang correction to the  $t=0$ value of
the free energy (term  proportional to $(na^3)^{1/2}$).  Starting from
expression (\ref{stringari_F})  for the free  energy \cite{Verney2015}
all the thermodynamic  functions entering Eq.~(\ref{stringari_c4}) are
easily  calculated using standard thermodynamic  relations.  The
normal  density (\ref{stringari_Landau})  is  instead evaluated  using
Eq.~(\ref{stringari_Landau}),
 written in terms  of the energy of the
elementary excitations.

The   idea    now   is   to    calculate   the   two    solutions   of
Eq.~(\ref{stringari_c4}) for a fixed value of $\tilde{t}$ of the order
of unity, taking the limit  $\eta \to 0$.  Physically this corresponds
to considering very  low temperatures (of the order of  $gn$) and  
small values of the gas parameter  $na^3$.  For example in the case of
$^{87}$Rb the  value of  the scattering length  is $a=100  a_0$ (where
$a_0$ is the Bohr radius) and typical values of the density correspond
to $na^3 \sim 10^{-6}$. This yields $\eta \approx 0.04$.

Writing  the solutions  of  Eq.~(\ref{stringari_c4}) in  terms of  the
$T=0$  Bogoliubov velocity  $c_B =  \sqrt{gn/m}$, one  finds that,  as
$\eta  \rightarrow 0$,  the two  sound velocities  only depend  on the
dimensionless parameter $\tilde{t}$ and are given by
\begin{align}
 c^2_+ &= c^2_B \; ,
&
 c^2_- &= c^2_B f(\tilde{t})
\label{stringari_c+-}
\end{align}
where   
$f(\tilde{t})   =   
\lim_{\eta\rightarrow  0}  
\frac{n}{n_n} \frac{T\bar{s}^2}{\bar{c}_v} \frac{1}{gn}
$.   
One actually  easily
finds  that  $\bar{s}   \propto  \eta^{3/2}$,  $\bar{c}_v  \propto
\eta^{3/2}$ and $n_n  \propto \eta^{3/2}$ as $\eta \to  0$ and that
$f$,  in  this  limit,  is consequently  a  function  of  $\tilde{t}$,
independent of $\eta$.
\begin{figure}
  \begin{minipage}{.47\textwidth}
    \centering
    \includegraphics[width=\textwidth]{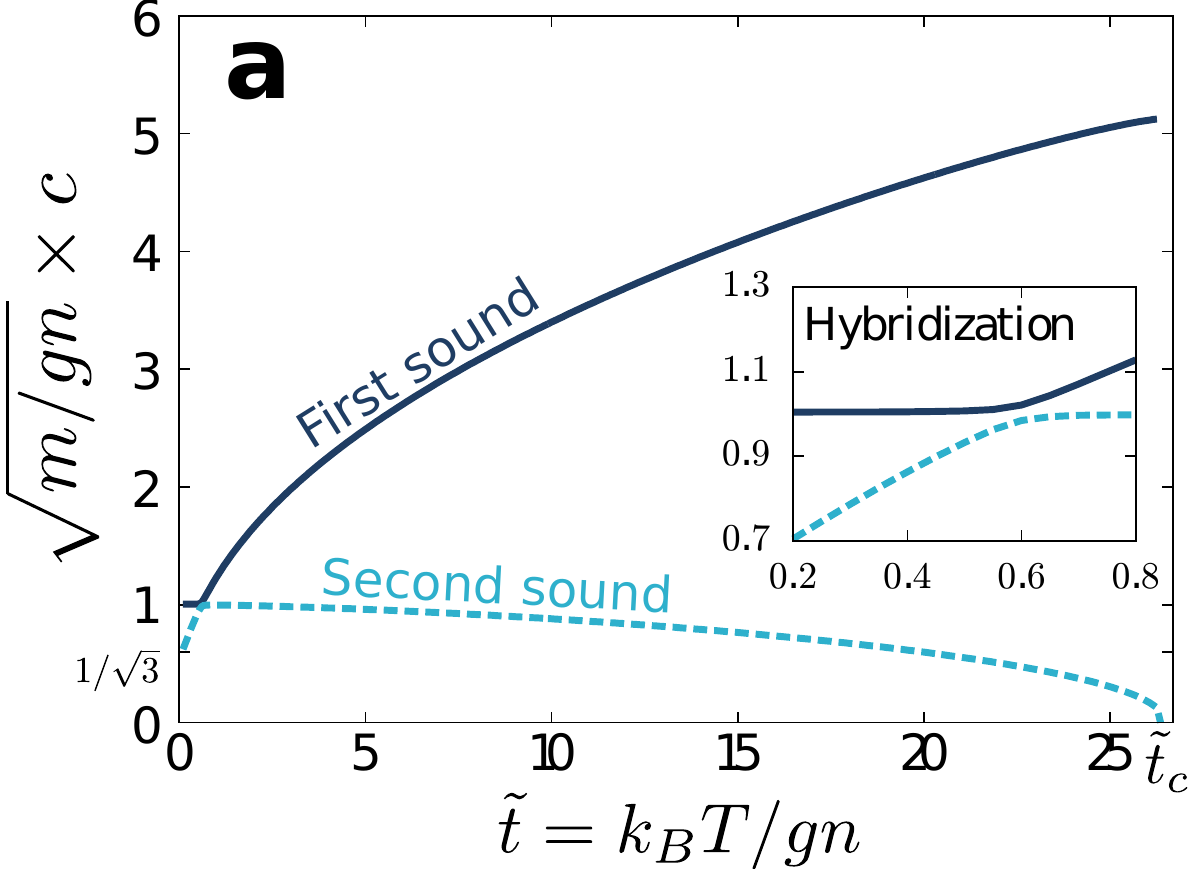}
  \end{minipage}
  \hspace*{.04\textwidth}
  \begin{minipage}{.47\textwidth}
    \centering
    \includegraphics[width=\textwidth]{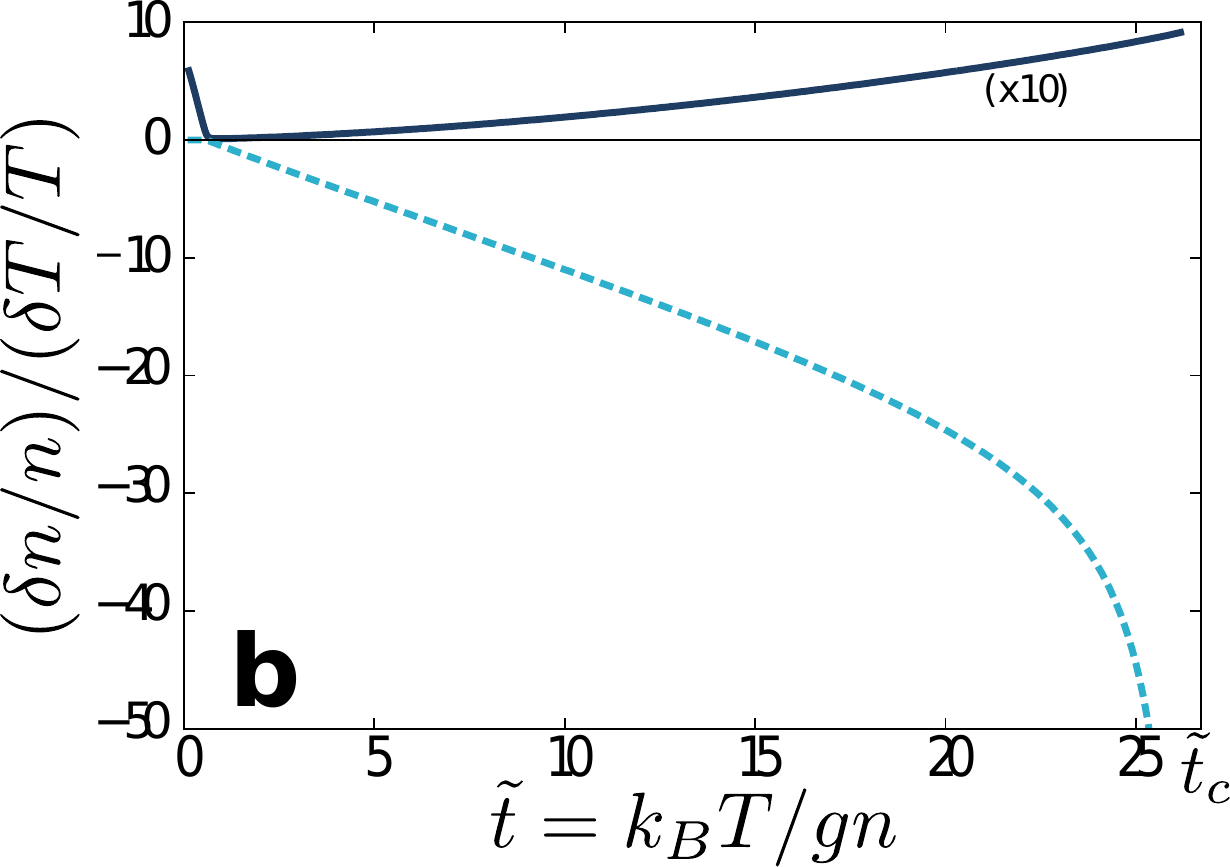}
  \end{minipage}
    \caption{ \label{stringari_speeds_deltandeltat}
\textbf{(a)}
Sound  velocities   (dashed light blue  and  solid dark  blue lines)   
computed  interpolating
Bogoliubov    theory     and    the    diagrammatic     approach    of
Ref.~\cite{Capogrosso2010},  over  the  whole  range  of  temperatures
$0<\tilde{t}<\tilde{t}_c$.   The inset  shows the
hybridization region. 
\textbf{(b)}
Ratio $(\delta n/n)/(\delta T/T)$ for the lower
branch (dashed light blue)  and the upper branch (solid dark  blue). 
The parameters
are chosen as in Fig.~\ref{stringari_speeds_deltandeltat}(a). 
For both graphs,  the   gas  parameter  is  chosen   to  be
$na^3=10^{-6}$, and the  critical point  
corresponds to  $\tilde{t}_c = T_c/gn= 26.7$.
From \cite{Verney2015}}
\end{figure}
In  the  $\eta \rightarrow  0$  limit,  the  two velocities  shown  in
Fig.~\ref{stringari_speeds_deltandeltat}(a)  
 cross   each    other   at   the   value
${\tilde{t}_{\mathrm{hyb}}\approx   0.6}$.   At   lower  temperatures   $c^2_-$
approaches,  as expected,  the zero-temperature  value $c^2_B/3$.   By
considering finite, although  small, values of $\eta$,  it is possible
to show that,  at the hybridization point, the two  branches exhibit a
gap  proportional to  $\eta^{3/4}$.  The  mechanism of  hybridization,
explicitly shown in the  inset of 
Fig.~\ref{stringari_speeds_deltandeltat}(a), permits
to  identify an  upper branch  $c_1$ (which  coincides with  $c_+$ for
$\tilde{t}  <  \tilde{t}_{\mathrm{hyb}}$  and   with  $c_-$  for  $\tilde{t}  >
\tilde{t}_{\mathrm{hyb}}$)  called  ``first  sound''. The  lower  branch  $c_2$
(called ``second sound'') instead  coincides with $c_-$ for $\tilde{t}
< \tilde{t}_{\mathrm{hyb}}$ and with $c_+$ for $\tilde{t} > \tilde{t}_{\mathrm{hyb}}$.

The  validity of  Eqs.~(\ref{stringari_c+-})  is limited  to very  low
temperatures  where the  thermal depletion  of the  condensate can  be
ignored  and  Bogoliubov  theory  can be  safely  applied.   When  the
temperature  is  comparable  to  the critical  temperature
$T_c$ (corresponding  to  $\tilde{t}_c=26.7$ in the  case of
Fig.~\ref{stringari_speeds_deltandeltat}(a),  where   we  have   chosen  $\eta=0.04$),
Bogoliubov  theory  is   no  longer  applicable.   In   fact  at  such
temperatures the thermal depletion of the condensate becomes important
and the  Bogoliubov expression for  the dispersion law  is inadequate.
The   superfluid    density   fraction,   calculated    according   to
Eq.~(\ref{stringari_Landau}) with  the $T=0$  value of  the Bogoliubov
dispersion law, vanishes  at $T\sim 1.2\,T_c$, well  above the critical
temperature, further  revealing the inadequacy of
the  theory  at  high  temperatures. As  anticipated  above  a  better
approach to be used at temperatures of the order of the critical value
is  the  diagrammatic  approach  of  Ref.~\cite{Capogrosso2010}  whose
predictions for the first and  second sound velocities at temperatures
higher   than   the   hybridization  temperature   are   reported   in
Fig.~\ref{stringari_speeds_deltandeltat}(a) and  which properly 
interpolates  with the
predictions of Bogoliubov theory near the hybridization point.

In  Fig.~\ref{stringari_speeds_deltandeltat}(b), 
we  show the  ratio
$(\delta  n/n)/(\delta  T/  T)$   between  the  relative  density  and
temperature  variations  calculated for  the  first  and second  sound
solutions    of   Eq.~(\ref{stringari_c4})    as    a   function    of
temperature. This quantity represents an important characterization of
the two  branches. It  is in fact  well known that  sound in  an ideal
classical gas is  an adiabatic oscillation characterized  by the value
$3/2$  for the  ratio $(\delta  n/n)/(\delta  T/ T)$.   For the  upper
solution of  the hydrodynamic equations  (first sound), this  ratio is
positive over the full range  of temperatures below the transition and
its  value  increases  with  $T$,  getting close  to  unity  near  the
transition.     The    most    important   feature    emerging    from
Fig.~\ref{stringari_speeds_deltandeltat}(b)  is that  the  ratio 
between  the
relative density  and temperature  changes associated with  the second
sound  solution, has  an opposite  sign and  a much  larger value  in
modulus, reflecting  that second  sound, for temperatures  larger than
the  hybridization  value is  dominated  by  the fluctuations  of  the
density, rather than  by the ones of the  temperature.  This important
feature  is also  revealed by  the  fact that,  in the  same range  of
temperatures,  second sound  practically exhausts  the compressibility
sum  rule  (\ref{stringari_kappaT1}).   This   result  can  be  easily
understood from Eq.~(\ref{stringari_Hu}).  In fact if $T  \gg gn$ one
has $mc^2_1 \gg (n\kappa_T)^{-1}$ and consequently $W_2 \gg W_1$.  The
fact that  the compressibility  sum rule  (\ref{stringari_kappaT1}) is
strongly affected by second sound is a remarkable feature exhibited by
the weakly-interacting  Bose gas  above the hybridization  point which
distinguishes    in  a  profound  way its behavior from  the one  of
strongly  interacting superfluids,  like $^4$He  or the unitary  Fermi 
gas, where the density fluctuations associated with second sound
are  very small.

From  an experimental  point of  view  the results  discussed in  this
section,   and  in   particular  
Fig.~\ref{stringari_speeds_deltandeltat}(b),
reveal that,  in dilute Bose gases,  second sound  is more  easily accessible  than first
sound, being  very sensitive to the coupling with
the density probe. This is  confirmed by the experimental identification
of second sound in the  experiment of Ref.~\cite{Meppelink2009}. It is
finally worth noticing  that the fact that second sound  in a Bose gas
can be easily  excited through a density probe is  not specific to the
3D case. Indeed, a similar behavior  takes place also in 2D Bose gases
\cite{Ozawa2014}  where its  measurement  could  provide an  efficient
determination of  the superfluid density, including  its discontinuity
at the Berezinski-Kosterlitz-Thouless transition (see Section 5).
 
\section{Second sound in the unitary Fermi gas} 

The  possibility  of determining  the  temperature  dependence of  the
superfluid density from  the measurement of second  sound represents a
major  challenge in  the  physics of  interacting  Fermi gases  where,
differently from the  case of weakly interacting  Bose gases discussed
in  the previous  section,  the
superfluid  density  $n_s$ cannot be identified with the condensate  
fraction and  the
theoretical determination of $n_s$  at finite temperature remains a
difficult problem from the many-body point of view. Recent progress in
the  experimental  measurement   of  second  sound  of   a  Fermi  gas
\cite{Sidorenkov2013} at unitarity has opened new perspectives in this
direction and the  first determination of the superfluid  density in a
Fermi superfluid.

First  predictions for  the temperature  dependence of  the first  and
second sound velocities  in the 3D unitary Fermi gas  were obtained in
\cite{Taylor2009} by calculating  the thermodynamic functions entering
the  two-fluid  Landau's  hydrodynamic  equation  (\ref{stringari_c4})
using      the      Nozieres     Schmitt-Rink      (NSR)      approach
\cite{Nozieres1985,Hu2006} (see Fig.~\ref{stringari_He4_TaylorFig}b).  
These
result have provided a first   estimate of the second
sound  velocity in  this  strongly interacting  Fermi  system.  It  is
remarkable to  see that the  qualitative behavior of the  second sound
velocity of  the unitary Fermi  gas looks very  similar to the  one of
superfluid $^4$He (Fig.  \ref{stringari_He4_TaylorFig}a).
  
In  the  following  we  will  discuss the  behavior  of  second  sound
developing  the hydrodynamic  formalism in  the presence  of a  highly
elongated  trap,  a  configuration   particularly  suited  to  explore
experimentally the  propagation of  sound.  The  effect of  the radial
confinement has the  important consequence that the  variations of the
temperature $\delta T(z,t)$ and of  the chemical potential $\delta \mu
(z,T)$,  as well  as  the  axial velocity  field  $v_n^z(z,t)$ do  not
exhibit any dependence on the radial coordinate during the propagation
of sound which can then be 
considered one  dimensional in nature, although  the local equilibrium
properties of  the system can  be described using the  3D themodynamic
functions in the local density  approximation.  The validity of the 1D
like  assumption is  ensured by  collisional effects  which restore  a
radial local thermodynamic equilibrium and requires the condition that
the  viscous  penetration  depth  
$\sqrt{\eta_s/mn_{n1}\omega}$  be
larger than the radial size of the system.  Here $n_{n_1}$ is the 1D
normal density, obtained by radial  integration of the normal density,
$\eta_s$ is the  shear viscosity and $\omega$ is the  frequency of the
sound wave.  In terms of  the radial trapping frequency $\omega_\perp$
the condition  can be written  in the form $\omega  \ll \omega^2_\perp
\tau$ where  $\tau$ is  a typical collisional  time here  assumed, for
simplicity, to characterize both the  effects of viscosity and thermal
conductivity.  The condition of radial local thermodynamic equilibrium
would have a  dramatic consequence in the presence of  a tube geometry
with hard  walls. In fact in  this case the viscosity  effect near the
wall would cause the vanishing of  the normal velocity field, with the
consequent  blocking  of  the  normal fluid.   The  resulting  motion,
involving only the superfluid fraction with the normal component at rest,
is called fourth  sound and was observed in liquid  helium confined in
narrow capillaries.  In the presence  of radial harmonic trapping this
effect  is absent  and the  normal  component can  propagate as  well,
allowing for the propagation of both first and second sound.

Under the above conditions of  local radial equilibrium one can easily
integrate       radially        the       hydrodynamic       equations
(\ref{stringari_dvsdt}-\ref{stringari_ds})
which then  keep the same
form as for the 3D uniform gas \cite{Bertaina2010}:
\begin{eqnarray}\label{stringari_Eq.1F1}
\partial_{t}n_1+\partial_z j_z =0
\end{eqnarray}
\begin{eqnarray}\label{stringari_Eq.2F1}
m\partial_t v_s^z=-\partial_z(\mu_1(z)+V_{ext}(z))
\end{eqnarray}
\begin{eqnarray}\label{stringari_Eq.3F1}
\partial_{t}s_1+\partial_z(s_1v_{n}^z)=0
\end{eqnarray}
\begin{eqnarray}
\partial_tj_z=
-\frac{\partial_z P_1}{m}
-\frac{n}{m}\partial_z V_{ext}(z) \; ,
\label{stringari_Eq.4F1}
\end{eqnarray}
where $n_1(z,t)=\int  dxdy\, n({\bf  r},t)$, 
$s_1(z,t)=\int  dxdy\, s({\bf r},t)$, and 
$P_1(z,t)=\int dxdy\, P({\bf  r},t)$ are the radial integrals
of their  3D counterparts,  namely the  particle density,  the entropy
density  and the  local  pressure. The  integration  accounts for  the
inhomogeneity  caused   by  the  radial  component   of  the  trapping
potential. In the above  equations $j_z = n_{n1}v_n^z +n_{s1}v_s^z$
is the current  density, $n_{s1}=\int dxdy\, n_s$  and 
$n_{n1}=\int dxdy\, n_n$ are the superfluid and  the normal 
1D densities respectively with
$n_1=n_{n1}+n_{s1}$,  while $v_s^z$  and $v_n^z$  are the  corresponding
velocity  fields. In  eqn (\ref{stringari_Eq.2F1})  $\mu_1(z,t) \equiv
\mu({\bf r}_\perp=0,z,t)$ is the chemical potential calculated on the
symmetry axis of the trapped gas. Its dependence on the 1D density and
on  the temperature  is  determined  by the  knowledge  of the  radial
profile which  can be  calculated employing the  equation of  state of
uniform matter in the local density approximation.

By  setting  the axial  trapping  $V_{ext}(z)$  equal to  zero,  which
corresponds to  considering a  cylindrical geometry,  we can  look for
sound  wave solutions  propagating with  a  phase factor  of the  form
$e^{i(qz-\omega  t)}$.  One  then derives  the same  Landau's equation
(\ref{stringari_c4})   obtained   in    uniform   matter,   with   the
thermodynamic functions replaced by  the corresponding 1D expressions,
in  particular  the  1D  entropy  per  particle  is  given  by  ${\bar
s}_1=s_1/n_1$ and the 1D specific  heat at constant pressure by ${\bar
c}_{p1} =T(\partial {\bar s}_1/\partial T)_{p_1}$.

Also  in the highly elongated 1D cigar configuration  the  Landau's equations  admit  two different  solutions,
corresponding  to   the  first   ($c_1$)  and  second   ($c_2$)  sound
velocities.   Simple results  for  the sound  velocities are  obtained
under the  condition (\ref{stringari_condition}), yielding the results 
\begin{align}
mc^2_1 &=\left(\frac{\partial P_1}{\partial n_1}\right)_{{\bar s}_1} 
&
mc^2_2 &=T\frac{n_{s1}{\bar s}^2_1}{n_{n1}{\bar c}_{p1} }
\label{stringari_c1dec}
\end{align}
for  the first and second  sound  velocity, respectively.  

Some    comments   are    in   order    here:   i)    The   assumption
(\ref{stringari_condition}),              yielding             results
(\ref{stringari_c1dec}) is  well  satisfied  in
strongly interacting  superfluid Fermi gases in  the whole temperature
interval below  $T_c$ due  to their  small compressibility  (from this
point  of view  the  behavior  of the  solutions  of the  hydrodynamic
equations deeply differ  from the case of dilute  Bose gases discussed
in the  previous Section).  ii) The  expression (\ref{stringari_c1dec})
for the second  sound velocity contains the specific  heat at constant
pressure.   This   reflects  the  fact  that   second  sound  actually
corresponds to a wave propagating  at constant pressure rather than at
constant  density.   This  difference   is  very  important  from  the
experimental  point of  view.   In fact,  even a  small  value of  the
thermal  expansion coefficient  is crucial  in order  to give  rise to
measurable density fluctuations during the propagation of second sound
\cite{Sidorenkov2013}.   iii)  The  thermodynamic  ingredients  $P_1$,
${\bar s}_1$ and  ${\bar c}_{P1}$ are known with good
precision at unitarity in a useful range of temperatures.  They can be
easily determined starting from the 3D thermodynamic relations holding
for  the uniform  Fermi gas  at unitarity.  

In the following  we will outline the calculation  of the themodynamic
functions in the  case of the unitary Fermi  gas \cite{Hou2013}.  At
unitarity, the only  length scales for uniform  configurations are the
inteparticle  distance, fixed  by  the  density of  the  gas, and  the
thermal wavelength $\lambda_T=\sqrt{2\pi \hbar^2/mT}$, fixed by the
temperature.   Correspondingly,   the  energy   scales  are   now  the
temperature and the Fermi temperature
\begin{equation}
T_F = \frac{\hbar^2}{2m} (3\pi^2n)^{2/3}
\label{stringari_TF}
\end{equation}
or, in  alternative, the chemical
potential $\mu$.  It  follows that at unitarity  all the thermodynamic
functions  can be  expressed  \cite{Ho2004} in  terms  of a  universal
function   $f_p(x)$   depending   on   the   dimensionless   parameter
$x\equiv\mu/T$. This function can  be defined
in terms of the pressure and the density of the gas, according to the relationships
\begin{align}
\label{stringari_Pn}
P\frac{\lambda_T^{3}}{T}&=f_{p}(x),
	&
n\lambda_T^{3} &= f_{p}'(x) \equiv f_n(x) \; .
\end{align}
From   Eq.~(\ref{stringari_Pn}) for the density  one derives   the   useful
relationship $T/T_{F}=4\pi/[3\pi^{2}f_{n}(x)]^{2/3}$
for the  ratio between  the temperature and  the Fermi  temperature in
terms  of  the  ratio  $x$.  

In terms  of $f_n$ and  $f_p$ we  can calculate all  the thermodynamic
functions  of   the  unitary  Fermi   gas.  For  example,   using  the
thermodynamic relation  $s/n=-(\partial \mu  /\partial T)_P$,  we find
the result $ \bar{s}=s/n=(5/2)f_{p}/f_{n}-x$ 
for the entropy per unit mass.
It  is worth  noting that  the above  equations for  the thermodynamic
functions of the unitary Fermi gas  are formally identical to the ones
of the  ideal Fermi gas,  the dimensionless functions $f_p$  and $f_n$
replacing, apart from a factor 2 caused by spin degeneracy, the usual
$F_{5/2}$      and      $F_{3/2}$       Fermi      integrals      (see
\cite{PitaevskiiStringari},  \S 16.1).  Since the entropy  $\bar{s}$    
depends  only on the
dimensionless     parameter    $x$,   from   Eqs.
(\ref{stringari_Pn})  one    finds that
during an  adiabatic transformation  the quantity  $P/n^{5/3}$ remains
constant, which is the same condition characterizing a non-interacting
monoatomic gas.

The scaling function $f_p(x)$ (and  hence its derivative $f_n(x)$) can
be determined through microscopic  many-body calculations or extracted
directly from  experiments carried out in  trapped configurations from
which it is possible to build  the equation of state of uniform matter
\cite{Ku2012}.   The   experimental  analysis  of   the  thermodynamic
functions  (in  particular  the  isothermal  compressibility  and  the
specific  heat) has  allowed for  the identification  of the  critical
temperature associated with the  superfluid phase transition for which
the authors of \cite{Ku2012} have found the result 
\begin{equation}
T_c = 0.167 T_F \; ,
\end{equation} 
in agreement  with the most recent  reliable theoretical calculations,
based  on  Quantum   Monte  Carlo  \cite{Burovski2006,Goulko2010}  and
diagrammatic  \cite{Haussmann2007}  techniques.  The  value  of  $T_c$
corresponds to $x_c=\mu_c/T_c= 2.48$.

\begin{figure}
  \centering
  \includegraphics[width=\textwidth]
  {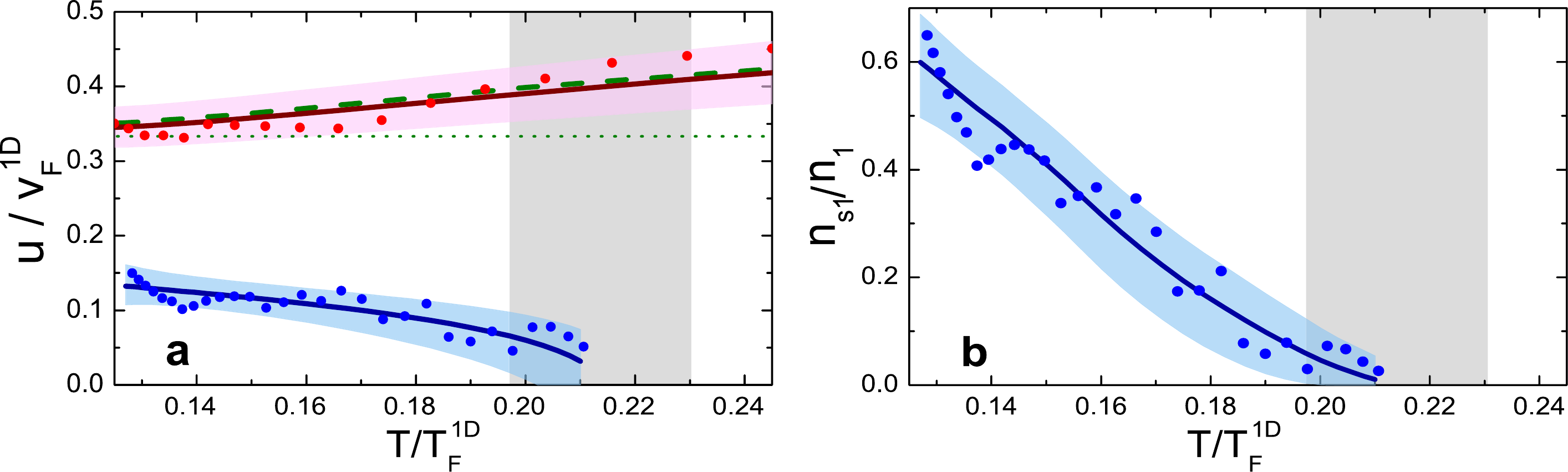}
  \caption{\label{stringari_fig:nature01}
\textbf{a}, Speeds of first and second  sound of the unitary Fermi gas
in a  highly elongated trap, normalized  to the local Fermi  speed and
plotted as a function of the reduced temperature.  The dashed curve is
a  prediction  based  on  Eq.~\ref{stringari_c11D} and  the  EOS  from
ref.~\cite{Ku2012}. The dotted horizontal line is the zero-temperature
limit  for   the  speed  of  first   sound.   \textbf{b},  Temperature
dependence   of  the   1D  superfluid   fraction  $n_{s1}/n_1$.   From
\cite{Sidorenkov2013}}
\end{figure}

Using  the local  density  approximation along  the radial  direction,
which   is  obtained   by  replacing   the  chemical   potential  with
$\mu=\mu_0-(1/2)m\omega^2_\perp   r^2_\perp$,   and  integrating   the
thermodynamic functions $P$, $n$ and  $s$ along the radial directions,
it  is  possible  to  construct  the  corresponding  1D  thermodynamic
functions in terms of the 1D chemical potential $\mu_1=x_1\,T$, which
are needed to calculate the ingredients entering the Landau's equation
for the 1D two sound velocities. One finds \cite{Hou2013}
\begin{align}
\label{stringari_P1n1}
P_1&= \frac{2\pi}{m\omega^2_\perp}\frac{T^2}{\lambda^3_T}f_q(x_1) \;,
	&
n_1(x_1,T)&= \frac{2\pi}{m\omega^2_\perp}\frac{T}{\lambda^3_T}f_p(x_1),
\end{align}
with $f_q=\int_{-\infty}^x dx'f_p(x')$,
allowing for the determination of the various thermodynamic functions. In particular 
the 1D entropy per unit mass takes the form $\bar{s}_1= s_1/n_1= (7/2)f_q(x_1)-x_1f_p(x_1)$. 
From the
above equations  it follows  immediately that  $(\partial P_1/\partial
n_1)_{\bar{s}_1}=(7/5)  P_1/n_1$  which  differs  from  the  adiabatic
result $(\partial P/\partial n)_{\bar{s}}=(5/3) P/n$ holding in the 3D
case.  Using Equation (\ref{stringari_c1dec}) one then predicts  the result  
\begin{equation}
mc^2_1=\frac{7}{5}\frac{P_1}{n_1}
\label{stringari_c11D}
\end{equation}
for the first  sound velocity
in  good   agreement  with   the  experimental findings  (see  Figure
\ref{stringari_fig:nature01}). 

Since  the temperature  dependence of  the superfluid  density is  not
known theoretically in the unitary Fermi gas with sufficient accuracy,
for  the  discussion  of  second  sound we  will  follow  a  different
strategy: we  will employ the  measured value  of the 1D  second sound
velocity   (see  discussion   below)  to   extract  $n_{s1}(T)$   from
Eq.  (\ref{stringari_c1dec}).   By the  way,  in  the low  temperature
regime, the calculation  of the thermodynamic functions  in the highly
elongated geometry predicts $c_2  \propto \sqrt{T} \to 0$, differently
from the the uniform 3D case  where $c_2\to c_1/\sqrt{3}$. Thus the 1D
second sound velocity $c_2$ tends to zero both as $T\to T_c$ and $T\to
0$.   As  a   consequence,  in   the  cigar   geometry,  the   condition
(\ref{stringari_condition})  is rather well satisfied  for all  
temperatures  and
Eq. (\ref{stringari_c1dec}) is expected to be particularly accurate.

Fig.~\ref{stringari_fig:nature01} shows the  measured sound velocities
in the  experiment of  \cite{Sidorenkov2013} carried  out in  a highly
elongated Fermi gas at unitarity.   The excitation of first and second
sounds  was  obtained  by  generating, respectively,  a  sudden  local
perturbation of  the density and of  the temperature in the  center of
the trapped  gas.  Due  to the  finite, although  small, value  of the
thermal  expansion  coefficient of  the  unitary  Fermi gas  also  the
thermal perturbation, generating the second  sound wave, gives rise to
a measurable  density pulse.   In this experiment  both the  first and
second sound velocities were obtained  by measuring, for a fixed value
of $T$, the  time dependence position of the  density pulses generated
by the perturbation.  These measurements give access to the dependence
of the sound velocity on the ratio $T/T_F^{1D}$ where $T_F^{1D}=
(15\pi/8)^{2/5}(\hbar\omega_\perp)^{4/5}
(\hbar^2 n_1^2/2m)^{1/5}$
is  a natural  definition for  Fermi temperature  in 1D  cylindrically
trapped  configurations  \cite{Sidorenkov2013,Hou2013}.  If  $n_1$  is
calculated  for an  ideal Fermi  gas at  zero temperature,  $T_F^{1D}$
coincides  with  the usual  3D  definition  of the  Fermi  temperature
(\ref{stringari_TF}), with $n$ calculated on the symmetry axis.  In the
presence  of  axial  trapping   the  value  of  $T/T_F^{1D}$  actually
increases  as  one moves  from  the  center,  because of  the  density
decrease,  and eventually  the  density pulse  reaches the  transition
point   where  the   superfluid  vanishes. 

\begin{figure}
  \centering
  \includegraphics[width=.5\textwidth]
  {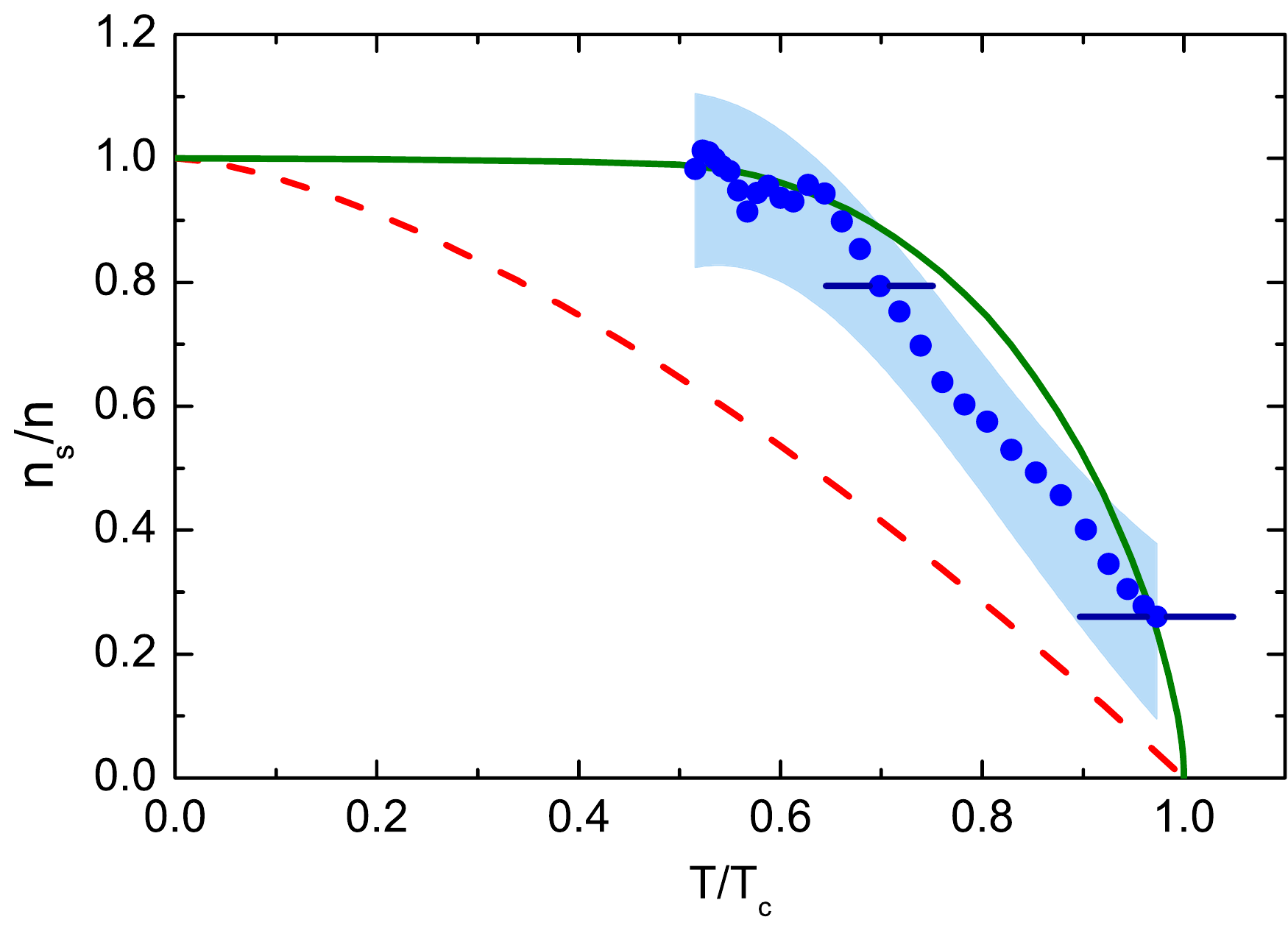}
  \caption{\label{stringari_fig:nature02}
Superfluid fraction for the homogeneous case. The data points
and  the  corresponding uncertainty  range  (shaded  region) show  the
superfluid  fraction for  a uniform  resonantly interacting  Fermi gas
versus   $T/T_c$  as   reconstructed  from   its  1D   counterpart  in
Fig.~\ref{stringari_fig:nature01}.  The  two   horizontal  error  bars
indicate  the  systematic  uncertainties resulting  from  the  limited
knowledge of the  critical temperature $T_c$. For  comparison, we show
the  fraction  for  superfluid  helium (solid  line)  as  measured  in
Ref.~\cite{Dash1957} and the textbook expression $1-(T/T_c)^{3/2}$ for
the Bose-Einstein  condensed fraction  of the  ideal Bose  gas (dashed
line). From \cite{Sidorenkov2013}}
\end{figure}

From the measurement of 
  the temperature dependence  of the 1D  superfluid density
  and  recalling   the  definition
$n_{s1}=\int dxdy\, n_s$, one can  reconstruct the 3D superfluid density
$n_s$ as a function of  the ratio $T/T_c$ \cite{Sidorenkov2013}, which
is reported in Fig.~\ref{stringari_fig:nature02}.
The  measurement of  $n_s$ reported  in the  figure represents  the
first experimental determination of  the temperature dependence of the
superfluid density  in a superfluid Fermi  gas and could be  used as a
benchmark for future many-body calculations.

\section{Second sound in the 2D Bose gas}

Two dimensional  superfluids differ  in a profound  way from  their 3D
counterparts.       In      fact,     the      Hohenberg-Mermin-Wagner
theorem~\cite{Hohenberg1967, Mermin1966}  rules out the  occurrence of
long range order  at finite temperature in 2D  systems with continuous
symmetry.   Furthermore, the  superfluid density  approaches a  finite
value  at   the  critical   point  of  a   2D  superfluid,   known  as
Berezinskii-Kosterlitz-Thouless                                  (BKT)
transition~\cite{Berezinskii1972,    Kosterlitz1972,   kosterlitz1973,
Nelson1977}, rather than vanishing, as happens in 3D.  With respect to
the  second  order  phase  transitions  characterizing  the  onset  of
superfluidity in 3D, the nature of  the BKT phase transition is deeply
different, being associated with the emergence of a topological order,
resulting from the  pairing of vortices and  antivortices.  A peculiar
property of these 2D systems is also the absence of discontinuities in
the  other   thermodynamic  functions  at  the   critical  temperature
characterizing the  transition to the  superfluid phase.  In  order to
identify  the  transition  point,  one  has  consequently  to  measure
suitable  transport  properties.   This  is the  case  of  the  recent
experiment  of Ref.~\cite{Desbuquois2012}  on  dilute two  dimensional
Bose gases  where the superfluid  critical velocity was measured  in a
useful range of  temperatures, pointing up the occurrence  of a sudden
jump at a critical temperature when one enters the superfluid regime.

In this Section we discuss the behavior of both first and second sound
in 2D superfluid gases, with  particular emphasis on the discontinuity
of their velocities  at the critical point, caused by  the jump of the
superfluid density.  Although the  thermodynamic behavior of 2D dilute
Bose      gases       is      now      well       understood      both
theoretically~\cite{Prokof'ev2001,   Prokof'ev2002,  Rancon2012}   and
experimentally~\cite{Hung2011,  Yefsah2011},  the measurement  of  the
superfluid  density  remains  one  of the  main open  issues.
Besides  the  prospect  of   measuring  the  superfluid  density,  the
measurement of second  sound itself would be  important because second
sound  has never  been measured  in any  2D system  so far.  In Helium
films, the  normal component of the  liquid is in fact  clamped to the
substrate and cannot participate in the propagation of sound, only the
superfluid being  free to  move (third sound).   In Helium  films, the
value of  $n_s$ became  accessible via  third sound~\cite{Rudnick1978}
and  torsional  oscillator measurements~\cite{Bishop1978},  confirming
the superfluid jump  at the transition.  Since the  trapping of dilute
atomic  gases  is  provided  by smooth  potentials,  second  sound  is
expected to propagate  in these systems also in 2D  and to be properly
described  by the  two  fluids Landau's  hydrodynamic equations.   Its
propagation in dilute Bose gases exhibits very peculiar
features  as  compared to  less  compressible  fluids like  helium  or
strongly  interacting  Fermi superfluid  gases. In  fact, in these  
latter systems second sound  can be
identified  as  an  entropy  oscillation  and  corresponds  with  good
accuracy to an  isobaric oscillation.  This is not the  case of dilute
Bose  gases which  are  highly compressible,  giving  rise to  sizable
coupling   effects   between   density   and   entropy   oscillations.
Furthermore,  with respect  to  the  3D case,  in  2D, the  superfluid
density  exhibits a  jump at  the transition  and this  shows up  as a
discontinuity   of   both   first    and   second   sound   velocities
\cite{Ozawa2014} as we will discuss in the following.

We  start our  investigation  by considering  the  Landau's two  fluid
hydrodynamic equations to  describe the dynamics of the  system in the
superfluid phase of a 2D uniform configuration. The third direction is
assumed to  be blocked  by a tight  harmonic confinement,  a condition
well achieved in current experiments.  We  will focus on the dilute 2D
Bose gas,  where all the  thermodynamic ingredients can be  written in
terms   of   dimensionless   functions~\cite{Prokof'ev2002,
Hung2011, Yefsah2011}.   These depend only  on the variable  
$x \equiv\mu_2/T$    
and     on    the    2D    coupling     constant    $g_2=
(\hbar^2/m)\sqrt{8\pi}a/l$, where $\mu$ is the chemical potential, $a$
is the three-dimensional scattering length,  and $l$ is the oscillator
length in the  confined direction.  Here, we assume $l  \gg a$ so that
the   interaction  is   momentum  independent~\cite{Petrov2000}.    We
introduce  the dimensionless  reduced pressure  $\mathcal{P}$ and  the
phase space density $\mathcal{D}$ by
\begin{align}
\label{stringari_P2n2}
	P_2\lambda_T^2/T &\equiv \mathcal{P}(x, mg_2/\hbar^2),
	&
	n_2\lambda_T^2 &\equiv\mathcal{D}(x,mg_2/\hbar^2),
\end{align}
where  $\lambda_T$  is the  thermal de  Broglie
wavelength, $P_2$ and  $n_2$ are the 2D pressure,  and particle number
density,     respectively.    The     simple    relation     $\partial
\mathcal{P}/\partial  x =  \mathcal{D}$  follows from  thermodynamics.
Starting from (\ref{stringari_P2n2}) one can evaluate the thermodynamic 
functions of the gas. In particular the entropy per unit mass, for a 
fixed value of $g_2$, depends solely on the parameter $x_2$:
$\bar{s}_2= s_2/n_2= 2\mathcal{P}/\mathcal{D}-x_2$.
The above equations reflect the  universal nature of the thermodynamic
behavior of  the 2D Bose gas,  for a fixed value  of the 
coupling  constant   $g_2$.   The  function  $\mathcal{D}$   has  been
numerically  calculated   for  values  of  $g_2$   much  smaller  than
$\hbar^2/m$  in~\cite{Prokof'ev2002},   and  both   $\mathcal{P}$  and
$\mathcal{D}$    have    been   theoretically~\cite{Rancon2012}    and
experimentally~\cite{Hung2011,   Yefsah2011}  determined   around  the
superfluid transition.   The results available from  different methods
well agree with each other.

The  superfluid density  $n_s$ cannot  be calculated  in terms  of the
universal  functions   introduced  above,  but  can   be  nevertheless
expressed  in terms  of  another  dimensionless function  $\lambda_T^2
n_s\equiv  \mathcal{D}_s  (x,g_2)$,  which   is  known  close  to  the
transition \cite{Prokof'ev2002}  as well  as in the  highly degenerate
phonon regime (large and positive $x$).  At the critical point one has
$\mathcal{D}_s=4$, which follows  from the universal Nelson-Kosterlitz
result $n_{2s} = 2mT_c/\pi\hbar^2$ \cite{Nelson1977},  
where $T_c$ is the BKT
transition  temperature,    providing an
important relationship between  the jump of the  superfluid density at
the transition and the value of the critical temperature. 

The superfluid transition of a  weakly interacting gas is predicted to
take place at the value ~\cite{Prokof'ev2001} 
$x_c =  (mg_2/\pi\hbar^2) \log(\xi_\mu/g_2)$
with $\xi_\mu \approx 13.2\hbar^2/m$.  
For example, for $g_2 = 0.1 \hbar^2/m$,
a value  relevant for  the experiments  of~\cite{Yefsah2011, Tung2010,
Desbuquois2012}, the critical point  corresponds to $x_c \approx 0.16$
or, in terms of the density, to 
$T_c = \{2\pi/\mathcal{D}(x_c)\}n_2/m \approx 0.76 n/m$.

\begin{figure}[htbp]
\begin{minipage}{.47\textwidth}
  \includegraphics[width=\textwidth]{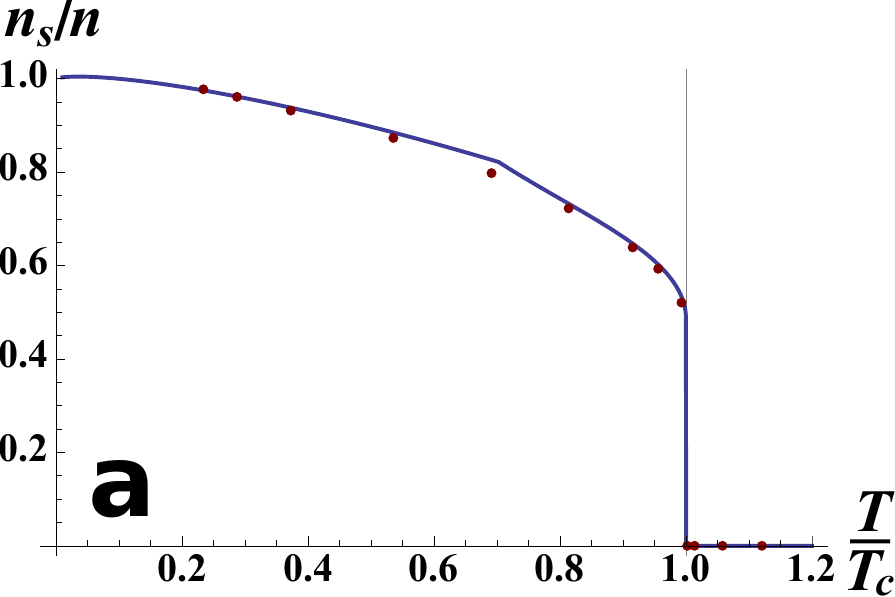}
\end{minipage}
\hspace*{.04\textwidth}
\begin{minipage}{.47\textwidth}
  \includegraphics[width=\textwidth]{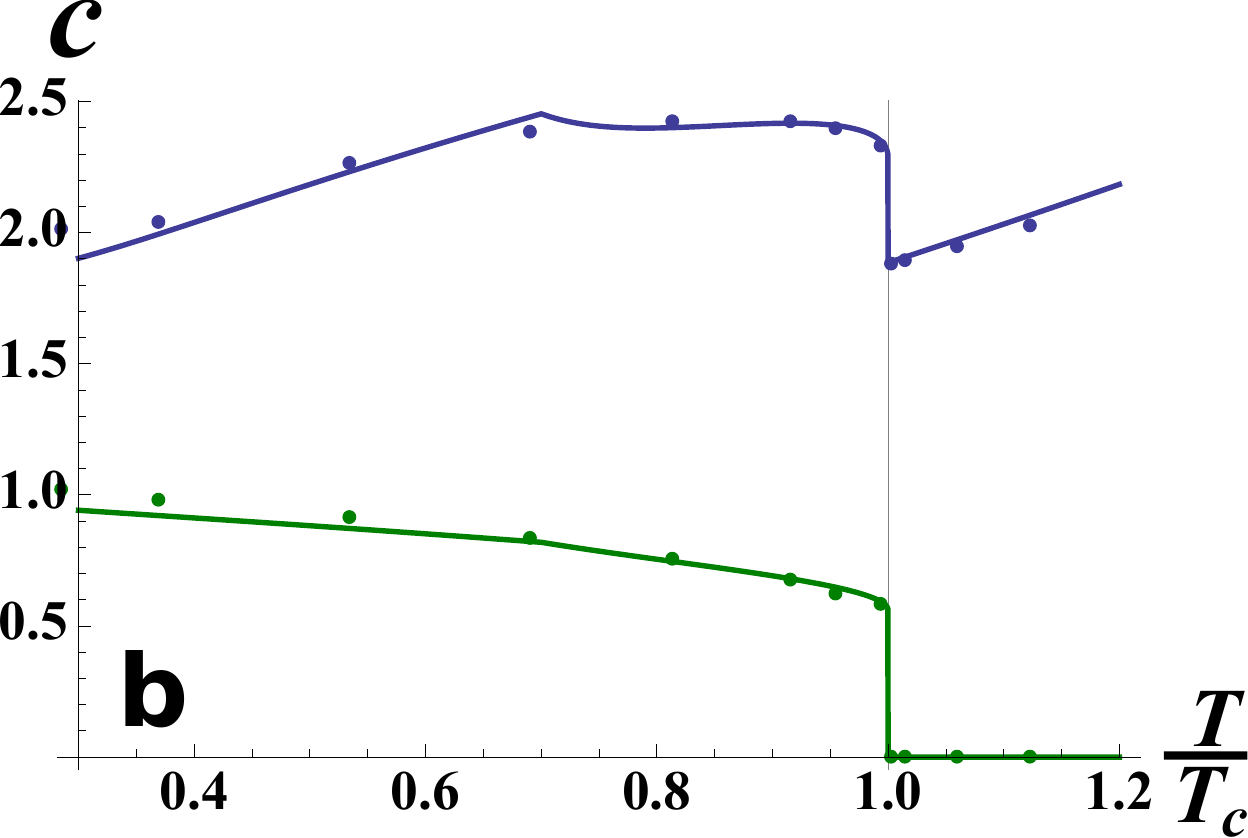}
\end{minipage}
\caption{  \label{stringari_nslines_c1c2}   
\textbf{(a)}
Normalized  superfluid  density
$n_s/n$  for $mg_2/\hbar^2  =  0.1$.  
The  line and  the  dots are  calculated,
respectively, from the approximate analytical and numerical results of
~\cite{Prokof'ev2002}.  Two analytical  expressions valid  at low  and
high temperatures  are connected  to give the  curve, resulting  in an
unphysical kink at $T/T_c \sim 0.7$. 
\textbf{(b)}
First  and second sound  velocities in
units  of  the  zero  temperature Bogoliubov  sound  velocity  $c_0  =
\sqrt{g_2n_2}/m$   with   $mg_2/\hbar^2=0.1$,   
calculated   by   solving
Eq.~(\ref{stringari_c4}).
From \cite{Ozawa2014}.}
\end{figure}

In  Fig.\ref{stringari_nslines_c1c2}(a)   
we  show  the   superfluid  fraction
$n_{2s}/n_2$, calculated as a function of $T/T_c$ for a fixed value of
the  total  density,  using  the data  from  \cite{Prokof'ev2002}.  The
figures correspond  to the value  $g_2 = 0.1\hbar^2/m$ and  points out
the large value  of the superfluid fraction at the  transition and the
consequent jump.   As we  will see  below the  jump of  the superfluid
density  as  well  as  the   large  value  of  the  thermal  expansion
coefficient play  an important role  to characterize the  solutions of
the Landau's equation near the transition.

In  Fig.~\ref{stringari_nslines_c1c2}(b) 
we  show the  values for  the first  and
second   sound    velocities   predicted    by   the    solutions   of
Eq.~(\ref{stringari_c4}). These  values are expressed in  units of the
zero temperature  value of the  Bogoliubov sound velocity  $c_0 \equiv
\sqrt{g_2n}/m$ and  are calculated at  fixed total density.   The most
remarkable  feature  emerging from  the  figure  is the  discontinuity
exhibited  by  both the  first  and  second  sound velocities  at  the
transition. Changing the value of the coupling constant $g_2$ does not 
affect the qualitative behavior of the sound velocities.

Using  the  parameters  from the  experiment  of~\cite{Desbuquois2012}
carried out on a gas of $^{87}$Rb  atoms 
($mg_2/\hbar^2= 0.093$, 
$n_2= 50/\mu\mathrm{m}^2$),     
we    predict     the    value     
$c_2    \approx 0.88~\mathrm{mm/s}$ 
for the second sound velocity at the transition.
This   value   is   close    to   the   critical   velocity   observed
in~\cite{Desbuquois2012},  thereby suggesting  that the  excitation of
second sound is  a possible mechanism for the onset  of dissipation in
this experiment.  

Concerning the physical characterization of the two sounds it is worth
recalling that in dilute Bose gases,  first and second sound cannot be
interpreted,  respectively, as  isoentropic and  isobaric oscillations
(see           Eqs.~(\ref{stringari_firstsoundvelocity})).  
In Section  III we have already shown  that in a
weakly interacting 3D Bose gas, the second sound speed $c_2$ is instead 
well
approximated     by     the    expression     (\ref{stringari_c-chiT})
~\cite{PitaevskiiStringari}  rather than  by 
Eq.~(\ref{stringari_firstsoundvelocity}),
in  the  relevant  region  of   temperatures  $T\gg  \mu$,  where  the
isoentropic  compressibility, the  entropy, and  the specific  heat at
constant volume  are very close to  the predictions of the  ideal Bose
gas model.   In two  dimensions Eq. (\ref{stringari_c-chiT}) describes  
exactly the  second sound velocity  the limit  of small
interactions ($g_2 \to 0$)  \cite{Ozawa2014}. For $g_2 = 0.1\hbar^2/m$
it is a good approximation (within $\sim 10\%$) to the second sound in
the whole  range of  temperatures shown  in 
Fig.~\ref{stringari_nslines_c1c2}(b).
On the other hand, the first  sound velocity around the transition can
be estimated  by solving Eq.~(\ref{stringari_c4}) for  small values of
$g_2$, which gives \cite{Ozawa2014}
\begin{align}
	c_1^2 = c_{10}^2 + \alpha T c_{20}^2
\end{align}
with       $c_{10}^2$       and      $c_{20}^2$       defined       by
Eqs. (\ref{stringari_firstsoundvelocity}),  with the  thermodynamic 
quantities  calculated in  2D.
This result shows that both  the nonzero thermal expansion coefficient
$\alpha$  and   the  discontinuity  in  the   superfluid  density  are
responsible for the jump of the  first sound velocity at the BKT 
transition.  

One finds that in the whole  interval of temperatures, second sound in
the  2D  Bose gas  corresponds  to  an  oscillation where  mainly  the
superfluid is moving,  the normal part remaining  practically at rest.
First  sound,  on  the  other  hand,  corresponds  to  an  oscillation
involving mainly the  normal component, similarly to the case of  3D Bose  
gases (see Section II).

In  order  to  measure  first  and second  sound,  a  first  important
requirement is the reachability of the collisional hydrodynamic regime
of  fast  collisions ($\omega  \tau  \ll  1$,  where $\omega$  is  the
frequency of  the sound and $\tau$  is a typical collisional  time) in
the  normal part.   This requirement  is likely  more problematic  for
first  sound due  to its  higher velocity.   The excitation  of second
sound should be  more easily accessible not only  because the velocity
is lower but also because in dilute  2D Bose gases it can be naturally
excited by density  perturbations.  For example, using  a sudden laser
perturbation,  applied to  the center  of the  trap, one  excites both
first  and second  sound  with  a relative  weight  given by  Equation
(\ref{stringari_Hu}).  Similarly  to the 3D  case also in the  2D Bose
gas one finds that in the  relevant temperature region 
$\mu \ll T < T_c$ 
one  has $mc^2_1  \gg (n_2\kappa_T)^{-1}$  and hence  
$W_2 \gg W_1$.  
In this range of  temperatures   second sound hence provides
most  of the  contribution  to the  compressibility  sum rule  thereby
making   its  experimental   excitation   favorable  through   density
perturbations.   At  lower temperatures,  in  the  phonon regime,  the
situation is modified  and a typical hybridization  effect between the
two sounds takes place \cite{PitaevskiiStringari}, as discussed in the
3D case in Section II.  As $T \to 0$, the thermodynamics is governed by
phonons   and  the   second  sound   velocity  approaches   the  value
$c_0/\sqrt{2}$.   Above   $T_c$,  the    difference   between  the
isoentropic  and isothermal  compressibilities is  instead responsible
for the occurrence of a diffusive mode at low frequency, 
like in 3D gases \cite{Hu2010}.

Typically,  in  experiments  of  dilute  ultracold  gases,  atoms  are
harmonically  trapped  also  along  the  radial  direction.   In  such
systems, $T/T_c$  depends on the  local density and thus,  by exciting
the sound modes through perturbing the  center of the trap and tracing
the propagation of the modes, one  could reveal the $T/T_c$ dependence 
of the
sound velocities, as observed in the case of three dimensional unitary
Fermi gas~\cite{Sidorenkov2013} (see the previous Section).

\section{Conclusions}

We have presented  an overview of recent  theoretical and experimental
advances in the study of the second sound velocity in ultracold atomic
gases.  Concerning  possible perspectives  and  open  problems in  the
future  studies  on  second  sound,   we  would  like  to  mention  the
experimental determination of the second  sound velocity in 2D quantum
gases  which would  allow  for the  determination  of the  temperature
dependence  of  the   superfluid  density  in  the   presence  of  the
Berzinski-Kosterlitz-Thouless  transition.   Other  topics   of  great
relevance,  and so  far  unexplored either  from  the theoretical  and
experimental perspective, are the study  of second sound in superfluid
mixtures of  Fermi and  Bose gases \cite{FerrierBarbut2014}  and the
role  of  spin-orbit  coupling,   causing  the  breaking  of  Galilean
invariance \cite{Zhu2012,Zheng2013,Ozawa2013}.

\section{Acknowledgments}

We  would  like  to  thank stimulating  collaborations  with  Gianluca
Bertaina, Rudolf  Grimm, Yanhua  Hou, Tomoki Ozawa,  Leonid Sidorenkov
and Meng  Khoon Tey. We  are also grateful  to David Papoular  for the
final preparation  of the paper. This  work has been supported  by ERC
through the QGBE grant.

\bibliography{stringaribib}

\begin{thebibliography}{66}
\expandafter\ifx\csname natexlab\endcsname\relax\def\natexlab#1{#1}\fi
\expandafter\ifx\csname selectlanguage\endcsname\relax
  \def\selectlanguage#1{\relax}\fi

\bibitem[\protect\citename{Abo{-}Shaeer {et~al.}, }2001]{AboShaeer2001}
Abo{-}Shaeer, J.~R., Raman, C., Vogels, J.~M., and Ketterle, W. 2001.
\newblock {\em Science}, {\bf 292}, 476.

\bibitem[\protect\citename{Abrikosov {et~al.}, }1975]{Abrikosov1975}
Abrikosov, A.~A., Gorkov, L.~P., and Dzyaloshinskii, I.~E. 1975.
\newblock {\em Methods of Quantum Field Theory in Statistical Physics}.
\newblock Dover.

\bibitem[\protect\citename{Albiez {et~al.}, }2005]{Albiez2005}
Albiez, M., Gati, R., F{\"o}lling, J., Hunsmann, S., Cristiani, M., and
  Oberthaler, M.~K. 2005.
\newblock {\em Phys. Rev. Lett.}, {\bf 95}, 010402.

\bibitem[\protect\citename{Arnold and Moore, }2001]{Arnold2001}
Arnold, P., and Moore, G. 2001.
\newblock {\em Phys. Rev. Lett.}, {\bf 87}, 120401.

\bibitem[\protect\citename{Baym {et~al.}, }1999]{Baym1999}
Baym, G., Blaizot, J.{-}P., Holzmann, M., Lalo{\"e}, F., and Vautherin, D.
  1999.
\newblock {\em Phys. Rev. Lett.}, {\bf 83}, 1703.

\bibitem[\protect\citename{Berezinskii, }1972]{Berezinskii1972}
Berezinskii, V.~L. 1972.
\newblock {\em Sov. Phys. JETP}, {\bf 34}, 610.

\bibitem[\protect\citename{Bertaina {et~al.}, }2010]{Bertaina2010}
Bertaina, G., Pitaevskii, L., and Stringari, S. 2010.
\newblock {\em Phys. Rev. Lett.}, {\bf 105}, 150402.

\bibitem[\protect\citename{Bishop and Reppy, }1978]{Bishop1978}
Bishop, D.~J., and Reppy, J.~D. 1978.
\newblock {\em Phys. Rev. Lett.}, {\bf 40}, 1727.

\bibitem[\protect\citename{Burovski {et~al.}, }2006]{Burovski2006}
Burovski, E., Prokof'ev, N., Svistunov, B., and Troyer, M. 2006.
\newblock {\em Phys. Rev. Lett.}, {\bf 96}, 160402.

\bibitem[\protect\citename{Capogrosso{-}Sansone {et~al.},
  }2010]{Capogrosso2010}
Capogrosso{-}Sansone, B., Giorgini, S., Pilati, S., Pollet, L., Prokof'ev, N.,
  Svistunov, B., and Troyer, M. 2010.
\newblock {\em New J. Physics}, {\bf 12}, 043010.

\bibitem[\protect\citename{Chin {et~al.}, }2004]{Chin2004}
Chin, C., Bartenstein, M., Altmeyer, A., Riedl, S., Jochim, S., {Hecker
  Denschlag}, J., and Grimm, R. 2004.
\newblock {\em Science}, {\bf 305}, 1128.

\bibitem[\protect\citename{Coddington {et~al.}, }2003]{Coddington2003}
Coddington, I., Engels, P., Schweikhard, V., and Cornell, E.~A. 2003.
\newblock {\em Phys. Rev. Lett.}, {\bf 91}, 100402.

\bibitem[\protect\citename{Dash and Taylor, }1957]{Dash1957}
Dash, J.~G., and Taylor, R.~D. 1957.
\newblock {\em Phys. Rev.}, {\bf 105}, 7.

\bibitem[\protect\citename{Desbuquois {et~al.}, }2012]{Desbuquois2012}
Desbuquois, R., Chomaz, L., Yefsah, T., L{\'e}onard, J., Beugnon, J.,
  Weitenberg, C., and Dalibard, J. 2012.
\newblock {\em Nature Phys.}, {\bf 8}, 645.

\bibitem[\protect\citename{Donnelly, }2009]{Donnelly2009}
Donnelly, R. 2009.
\newblock {\em Physics Today}, {\bf 62}(10), 34.

\bibitem[\protect\citename{Ferrier{-}Barbut {et~al.}, }2014]{FerrierBarbut2014}
Ferrier{-}Barbut, I., Delehaye, M., Laurent, S., Grier, A.~T., Pierce, M., Rem,
  B.~S., Chevy, F., and Salomon, C. 2014.
\newblock {\em Science}, {\bf 345}, 1035.

\bibitem[\protect\citename{Goulko and Wingate, }2010]{Goulko2010}
Goulko, O., and Wingate, M. 2010.
\newblock {\em Phys. Rev. A}, {\bf 82}, 053621.

\bibitem[\protect\citename{Griffin {et~al.}, }2009]{GriffinNikuniZaremba2009}
Griffin, A., Nikuni, T., and Zaremba, E. 2009.
\newblock {\em Bose{-}Condensed Gases at Finite Temperatures}.
\newblock Cambridge University Press (New York).

\bibitem[\protect\citename{Gu{\'e}ry{-}Odelin and Stringari,
  }1999]{GueryOdelin1999}
Gu{\'e}ry{-}Odelin, D., and Stringari, S. 1999.
\newblock {\em Phys. Rev. Lett.}, {\bf 83}, 4452.

\bibitem[\protect\citename{Haussmann {et~al.}, }2007]{Haussmann2007}
Haussmann, R., Rantner, W., Cerrito, S., and Zwerger, W. 2007.
\newblock {\em Phys. Rev. A}, {\bf 75}, 023610.

\bibitem[\protect\citename{Ho, }2004]{Ho2004}
Ho, {T.}-{L.} 2004.
\newblock {\em Phys. Rev. Lett.}, {\bf 92}, 090402.

\bibitem[\protect\citename{Hohenberg, }1967]{Hohenberg1967}
Hohenberg, P.~C. 1967.
\newblock {\em Phys. Rev.}, {\bf 158}, 383.

\bibitem[\protect\citename{Hou {et~al.}, }2013]{Hou2013}
Hou, {Y.-H.}, Pitaevskii, L.~P., and Stringari, S. 2013.
\newblock {\em Phys. Rev. A}, {\bf 88}, 043630.

\bibitem[\protect\citename{Hu {et~al.}, }2006]{Hu2006}
Hu, H., Liu, X.{-}{J.}, and Drummond, P.~D. 2006.
\newblock {\em Phys. Rev. A}, {\bf 73}, 023617.

\bibitem[\protect\citename{Hu {et~al.}, }2010]{Hu2010}
Hu, H., Taylor, E., Liu, {X.-J.}, Stringari, S., and Griffin, A. 2010.
\newblock {\em New J. Phys.}, {\bf 12}, 043040.

\bibitem[\protect\citename{Hung {et~al.}, }2011]{Hung2011}
Hung, {C.-L.}, Zhang, X., Gemelke, N., and Chin, C. 2011.
\newblock {\em Nature (London)}, {\bf 470}, 236.

\bibitem[\protect\citename{Kashurnikov {et~al.}, }2001]{Kashurnikov2001}
Kashurnikov, V.~A., Prokof'ev, N.~V., and Svistunov, B.~V. 2001.
\newblock {\em Phys. Rev. Lett.}, {\bf 87}, 120402.

\bibitem[\protect\citename{Kosterlitz and Thouless, }1972]{Kosterlitz1972}
Kosterlitz, J.~M., and Thouless, D.~J. 1972.
\newblock {\em J. Phys. C}, {\bf 5}, L124.

\bibitem[\protect\citename{Kosterlitz and Thouless, }1973]{kosterlitz1973}
Kosterlitz, J.~M., and Thouless, D.~J. 1973.
\newblock {\em J. Phys. C.}, {\bf 6}, 1181.

\bibitem[\protect\citename{Ku {et~al.}, }2012]{Ku2012}
Ku, M. J.~H., Sommer, A.~T., Cheuk, L.~W., and Zwierlein, M.~W. 2012.
\newblock {\em Science}, {\bf 335}, 563.

\bibitem[\protect\citename{Landau, }1941]{Landau1941}
Landau, L.~D. 1941.
\newblock {\em J. Phys. USSR}, {\bf 5}, 71.

\bibitem[\protect\citename{Landau and Lifshitz, }1987]{LandauLifshitz1987}
Landau, L.~D., and Lifshitz, E.~M. 1987.
\newblock {\em Fluid Mechanics}.
\newblock Pergamon (Oxford).

\bibitem[\protect\citename{Lee and Yang, }1959]{LeeYang1959}
Lee, T.~D., and Yang, C.~N. 1959.
\newblock {\em Phys. Rev.}, {\bf 113}, 1406.

\bibitem[\protect\citename{Lifshitz, }1944]{Lifshitz1944}
Lifshitz, E.M. 1944.
\newblock {\em J. Phys. USSR}, {\bf 8}, 110.

\bibitem[\protect\citename{Madison {et~al.}, }2000]{Madison2000}
Madison, K.~W., Chevy, F., Wohlleben, W., and Dalibard, J. 2000.
\newblock {\em Phys. Rev. Lett.}, {\bf 84}, 806.

\bibitem[\protect\citename{Marag{\`o} {et~al.}, }2000]{Marago2000}
Marag{\`o}, O.~M., Hopkins, S.~A., Arlt, J., Hodby, E., Hechenblaikner, G., and
  Foot, C.~J. 2000.
\newblock {\em Phys. Rev. Lett.}, {\bf 84}, 2056.

\bibitem[\protect\citename{Matthews {et~al.}, }1999]{Matthews1999}
Matthews, M.~R., Anderson, B.~P., Haljan, P.~C., Hall, D.~S., Wieman, C.~E.,
  and Cornell, E.~A. 1999.
\newblock {\em Phys. Rev. Lett.}, {\bf 83}, 2498.

\bibitem[\protect\citename{Meppelink {et~al.}, }2009]{Meppelink2009}
Meppelink, R., Koller, S.~B., and van~der Straten, P. 2009.
\newblock {\em Phys. Rev. A}, {\bf 80}, 043605.

\bibitem[\protect\citename{Mermin and Wagner, }1966]{Mermin1966}
Mermin, N.~D., and Wagner, H. 1966.
\newblock {\em Phys. Rev. Lett.}, {\bf 17}, 1133.

\bibitem[\protect\citename{Miller {et~al.}, }2007]{Miller2007}
Miller, D.~E., Chin, J.~K., Stan, C.~A., Liu, Y., Setiawan, W., Sanner, C., and
  Ketterle, W. 2007.
\newblock {\em Phys. Rev. Lett.}, {\bf 99}, 070402.

\bibitem[\protect\citename{Nelson and Kosterlitz, }1977]{Nelson1977}
Nelson, D.~R., and Kosterlitz, J.~M. 1977.
\newblock {\em Phys. Rev. Lett.}, {\bf 39}, 1201.

\bibitem[\protect\citename{Nozi{\`e}res and Schmitt{-}Rink,
  }1985]{Nozieres1985}
Nozi{\`e}res, P., and Schmitt{-}Rink, S. 1985.
\newblock {\em J. Low Temp. Phys.}, {\bf 59}, 195.

\bibitem[\protect\citename{Onofrio {et~al.}, }2000]{Onofrio2000}
Onofrio, R., Raman, C., Vogels, J.~M., Abo{-}Shaeer, J.~R., Chikkatur, A.~P.,
  and Ketterle, W. 2000.
\newblock {\em Phys. Rev. Lett.}, {\bf 85}, 2228.

\bibitem[\protect\citename{Ozawa and Stringari, }2014]{Ozawa2014}
Ozawa, T., and Stringari, S. 2014.
\newblock {\em Phys. Rev. Lett.}, {\bf 112}, 025302.

\bibitem[\protect\citename{Ozawa {et~al.}, }2013]{Ozawa2013}
Ozawa, T., Pitaevskii, L.~P., and Stringari, S. 2013.
\newblock {\em Phys. Rev. A}, {\bf 87}, 063610.

\bibitem[\protect\citename{Peshkov, }1946]{Peshkov1946}
Peshkov, V.~P. 1946.
\newblock {\em J. Phys. USSR}, {\bf 10}, 389.

\bibitem[\protect\citename{Petrov {et~al.}, }2000]{Petrov2000}
Petrov, D.~S., Holzmann, M., and Shlyapnikov, G.~V. 2000.
\newblock {\em Phys. Rev. Lett.}, {\bf 84}, 2551.

\bibitem[\protect\citename{Pines and Nozi{\`e}res, }1990]{NozieresPines}
Pines, D., and Nozi{\`e}res, P. 1990.
\newblock {\em Theory of Quantum Liquids}.
\newblock Addison-Wesley (Redwood City).

\bibitem[\protect\citename{Pitaevskii and Stringari,
  }2016]{PitaevskiiStringari}
Pitaevskii, L.~P., and Stringari, S. 2016.
\newblock {\em Bose{-}Einstein Condensation and Superfluidity}.
\newblock Oxford University Press (New York).

\bibitem[\protect\citename{Prokof'ev and Svistunov, }2002]{Prokof'ev2002}
Prokof'ev, N., and Svistunov, B. 2002.
\newblock {\em Phys. Rev. A}, {\bf 66}, 043608.

\bibitem[\protect\citename{Prokof'ev {et~al.}, }2001]{Prokof'ev2001}
Prokof'ev, N., Ruebenacker, O., and Svistunov, B. 2001.
\newblock {\em Phys. Rev. Lett.}, {\bf 87}, 270402.

\bibitem[\protect\citename{Ran{\c c}on and Dupuis, }2012]{Rancon2012}
Ran{\c c}on, A., and Dupuis, N. 2012.
\newblock {\em Phys. Rev. A}, {\bf 85}, 063607.

\bibitem[\protect\citename{Riedl {et~al.}, }2011]{Riedl2011}
Riedl, S., {S{\'a}nchez Guajardo}, E.~R., Kohstall, C., {Hecker Denschlag}, J.,
  and Grimm, R. 2011.
\newblock {\em New J. Phys.}, {\bf 13}, 035003.

\bibitem[\protect\citename{Rudnick, }1978]{Rudnick1978}
Rudnick, I. 1978.
\newblock {\em Phys. Rev. Lett.}, {\bf 40}, 1454.

\bibitem[\protect\citename{Sidorenkov {et~al.}, }2013]{Sidorenkov2013}
Sidorenkov, L.~A., Tey, M.~K., Grimm, R., Hou, {Y.-H.}, Pitaevskii, L., and
  Stringari, S. 2013.
\newblock {\em Nature (London)}, {\bf 498}, 78.

\bibitem[\protect\citename{Stringari, }1996]{Stringari1996}
Stringari, S. 1996.
\newblock {\em Phys. Rev. Lett.}, {\bf 77}, 2360.

\bibitem[\protect\citename{Taylor {et~al.}, }2009]{Taylor2009}
Taylor, E., Hu, H., Liu, {X.-J.}, Pitaevskii, L.~P., Griffin, A., and
  Stringari, S. 2009.
\newblock {\em Phys. Rev. A}, {\bf 80}, 053601.

\bibitem[\protect\citename{Tisza, }1940]{Tisza1940}
Tisza, L. 1940.
\newblock {\em J. Phys. Radium}, {\bf 1}, 164.

\bibitem[\protect\citename{Tung {et~al.}, }2010]{Tung2010}
Tung, S., Lamporesi, G., Lobser, D., Xia, L., and Cornell, E.~A. 2010.
\newblock {\em Phys. Rev. Lett.}, {\bf 105}, 230408.

\bibitem[\protect\citename{Verney {et~al.}, }2015]{Verney2015}
Verney, L., Pitaevskii, L., and Stringari, S. 2015.
\newblock {\em arXiv:1506.06690}.

\bibitem[\protect\citename{Yefsah {et~al.}, }2011]{Yefsah2011}
Yefsah, T., Desbuquois, R., Chomaz, L., G{\"u}nther, K.~J., and Dalibard, J.
  2011.
\newblock {\em Phys. Rev. Lett.}, {\bf 107}, 130401.

\bibitem[\protect\citename{Zaremba, }1998]{Zaremba1998}
Zaremba, E. 1998.
\newblock {\em Phys. Rev. A}, {\bf 57}, 518.

\bibitem[\protect\citename{Zaremba {et~al.}, }1999]{Zaremba1999}
Zaremba, E., Nikuni, T., and Griffin, A. 1999.
\newblock {\em J. Low Temp. Phys.}, {\bf 116}, 277.

\bibitem[\protect\citename{Zheng {et~al.}, }2013]{Zheng2013}
Zheng, W., Yu, {Z.}-{Q.}, Cui, X., and Zhai, H. 2013.
\newblock {\em J. Phys. B}, {\bf 46}, 134007.

\bibitem[\protect\citename{Zhu {et~al.}, }2012]{Zhu2012}
Zhu, Q., Zhang, C., and Wu, B. 2012.
\newblock {\em Europhys. Lett.}, {\bf 100}, 50003.

\bibitem[\protect\citename{Zwierlein {et~al.}, }2005]{Zwierlein2005}
Zwierlein, M.~W., Abo{-}Shaeer, J.~R., Schirotzek, A., Schunck, C.~H., and
  Ketterle, W. 2005.
\newblock {\em Nature}, {\bf 435}, 1047.

\end{thebibliography}
\bibliographystyle{cambridgeauthordate}

\end{document}